\documentclass[journal=jpccck,manuscript=article]{achemso}
\usepackage[version=3]{mhchem} 
\usepackage{graphicx}
\usepackage{epsfig}
\usepackage{multirow}
\usepackage{dcolumn}
\usepackage{bm}
\usepackage{subfigure}
\usepackage{color}
\usepackage{amsmath}
\usepackage{amssymb}
\usepackage{amsthm}
\usepackage{graphicx}
\usepackage[version=3]{mhchem}
\usepackage{dcolumn}
\usepackage{multirow}
\usepackage{bm}
\usepackage{subfigure}
\usepackage{todonotes}
\mciteErrorOnUnknownfalse

\title[An \textsf{achemso} demo]
{Lattice instability, anharmonicity and Raman spectra of BaO under high pressure: A first principles study}
\author{K. Lavanya}
\affiliation{Jawaharlal Nehru Technological University (JNTU), Hyderabad, 500085, Telangana, India.}
\alsoaffiliation{Telangana Social Welfare Residential Degree College for Women (TSWRDC-W), Nizamabad, 503002, Telangana, India.}
\author{N. Yedukondalu}
\email{nykondalu@gmail.com}
\affiliation{Department of Geosciences, State University of New York, Stony Brook, New York 11794-2100, USA}
\author{S. C. Rakesh Roshan}
\affiliation{Rajiv Gandhi University of Knowledge Technologies, Basar, Telangana-504107, India}
\alsoaffiliation{Department of Physics, National Institute of Technology-Warangal, Telangana, India}
\author{Shweta D. Dabhi}
\affiliation{Department of Physical Sciences, P D Patel Institute of Applied Sciences, Charotar University of Science and Technology, CHARUSAT Campus, Changa, 388421, Gujarat, India}
\author{Suresh Sripada}
\affiliation{Department of Physics, JNTUHCEJ, 505501 Karimnagar, Telangana, India}
\author{M.Sainath}
\email{sai1968@gmail.com}
\affiliation{Department of Physics, IcfaiTech, IFHE, Hyderabad 501203, Telangana, India.}
\author{Lars Ehm}
\affiliation{Department of Geosciences, State University of New York, Stony Brook, New York 11794-2100, USA}
\author{John B. Parise}
\affiliation{Department of Geosciences, State University of New York, Stony Brook, New York 11794-2100, USA}

\begin{document}
\begin{abstract}
Alkaline-earth metal oxides, in particular MgO and CaO dominate Earth’s lower mantle, therefore, exploring high pressure behavior of this class of compounds is of significant geophysical research interest.  Among the alkaline-earth metal oxides, BaO exhibits rich polymorphism in the pressure range of 0-1.5 Mbar. Static enthalpy calculations revealed that BaO undergoes a pressure induced structural phase transition from NaCl-type (B1) $\rightarrow$  NiAs-type (B8) $\rightarrow$ distorted CsCl-type (d-B2) $\rightarrow$ CsCl-type (B2) at 5.1, 19.5, 120 GPa respectively. B1 $\rightarrow$ B8 $\&$ B8 $\rightarrow$ d-B2 transitions are found to be first order in nature whereas d-B2$\rightarrow$ B2 is a second order or weak first order phase transition with displacive nature. Interestingly, d-B2 phase shows stability over a wide pressure range of $\sim$19.5-113 GPa. Mechanical and dynamical stabilities of ambient and high pressure phases are demonstrated through the computed second order elastic constants and phonon dispersion curves, respectively. Under high pressure,  significant phonon softening and soft phonon mode along M-direction are observed for B8, d-B2 and B2 phases, respectively. Pressure dependent Raman spectra suggests a phase transition from d-B2 to Raman inactive phase under high pressure. Overall, the present study provides a comprehensive understanding of lattice dynamics and underlying mechanisms behind pressure induced structural phase transitions in BaO. 
\end{abstract}
\maketitle
\section{Introduction}

Alkaline-earth metal oxides, MO (M = Mg, Ca, Sr, Ba) have gained a considerable attention due to their technological applications in diverse fields,\cite{Cho1999,Speier1997,Duan2010,Medeiros2007,Zannotti2015} ranging from catalysis \cite{app-1-Prinetto2003,Choudhary1998,Au1998} to electrode coating \cite{app2-Mishra2004} and interfaces \cite{McKee2001}. These metal oxides are suitable candidates for CO$_2$ sorbent applications due to their high absorption capacity at moderate temperatures.\cite{Duan2010} Moreover, these mono-oxides are prime candidates for the Earth's interior, especially light metal oxides such as MgO and CaO are abundant in the Earth's lower mantle, in which the pressure varies from 24 to 136 GPa from top of the upper mantle to the lower mantle.\cite{Chen1998,mantle-Margrave1979,Bovolo2005,anderson1989theory,McDonough1995} Therefore, exploring high pressure and/or temperature behavior of these metal oxides  provides an insight on the deep interior of the Earth. These metal oxides crystallize in the rocksalt NaCl-type (B1) structure with space group $Fm\bar{3}m$ at ambient conditions. Upon compression, these metal oxides usually undergo a body centred cubic CsCl-type (B2) structure and it is well established for MgO, CaO and SrO. Amongst all these compounds, BaO has interesting structural properties and it exhibits a rich polymorphism in the pressure range of 0-1.5 Mbar over the remaining alkaline earth oxides, CaO,\cite{Aguado2003,Alfredsson2005,Cinthia2015,Pozhivatenko2020,Kunduru2021,Krk2018,Kholiya2010,Jeanloz1979,Bhardwaj2010} SrO.\cite{Aguado2003,Alfredsson2005,Cinthia2015,Pozhivatenko2020,Hou2021,SrO_Lukaevi2011,AbdusSalam2019,Souadkia2012,Sato1981} Therefore, understanding the high pressure structural and lattice dynamical behavior of BaO is crucial as this can be a possible candidate of the Earth's mantle\cite{mantle-Margrave1979}. Several high pressure X-ray diffraction (HP-XRD) measurements\cite{Liu1972,Sato1981,Jeanloz1979,Weir1986,Rieder1973,Galtier1972,Miyanishi2015,Root2015} and theoretical studies using first principles calculations were focused on exploring the structural phase transitions\cite{Duan2008,Rajput2020,Bouchet2019,AbdusSalam2019,Cinthia2015,Krk2018,Hou2021,Yu2019}, metallization \cite{Ghebouli2010,Baltache2004,Baumeier2007,Beiranvand2021,Sobolev2016,Kunduru2021,Kumar2013,Sun2021}, elastic properties \cite{Tsuchiya2001,Guo2006,Pandey2010,Singh2022}, thermodynamic stability \cite{Hou2021,Rajput2021,Lukaevi2011-BaO} of MO (M = Mg, Ca, Sr, Ba) compounds under high pressure. Similarly, efforts have been put forward by the researchers to explore the phase diagram of BaO under high pressure from both experimental and theoretical perspective. A HP-XRD study\cite{Liu1972} revealed that BaO undergoes a structural transition from B1 to distorted B2 (d-B2) at 9.2 GPa and remains in the same structure till 29 GPa. Energy dispersive X-ray diffraction study \cite{Weir1986} disclosed that BaO transforms from B1 to an intermediate hexagonal NiAs-type (B8) structure at around 10 GPa, later to a tetragonal PH$_4$I-type (d-B2) structure with pseudo 8-fold coordination at around 15 GPa and this phase approaches to B2 structure with increasing pressure above 1 Mbar. First principles studies\cite{Uludoan2001, Alfredsson2005, Amorim2006} reported that BaO transforms from B1 to B8 phase at 11.3 GPa,\cite{Uludoan2001} 5 GPa,\cite{Alfredsson2005} 4.3 GPa\cite{Amorim2006} then to d-B2 from B8 at 21.5 GPa,\cite{Uludoan2001} 13 GPa,\cite{Alfredsson2005} 23.2 GPa,\cite{Amorim2006} upon further compression, the d-B2 phase transforms to B2 phase at 62 GPa,\cite{Uludoan2001} 50 GPa,\cite{Alfredsson2005} 108 GPa.\cite{Amorim2006} Pandey et al \cite{Pandey2010} and Jog et al \cite{Jog1985} suggested a direct B1 to B2 phase transition at 100 and 86 GPa, respectively without any intermediate high pressure phases.

On the other hand, lattice dynamics of B1 phase is well studied at ambient as well as at high pressure using first principles calculations\cite{wu2002applications,Aguado2003, Musari2018,Rajput2021} and inelastic neutron scattering technique.\cite{Chang1975} However, very limited studies\cite{Lukaevi2011-BaO} are available exploring lattice dynamics of high pressure phases, underlying mechanisms behind the observed structural phase transitions and spectroscopic properties under high pressure in BaO. Therefore, in the present study, we have systematically investigated the structural phase transitions, lattice dynamics of different polymorphs and Raman spectra of B8 and d-B2 phases of BaO under high pressure.

\section{Computational details}
All the first principles calculations were carried out using projected augmented wave (PAW) method as implemented in Vienna ab-initio simulation package (VASP)\cite{Kresse1996}. Electron-electron interactions are treated within generalized gradient approximation (GGA) with Perdew–Burke-Ernzerhof (PBE) functional while PAW pseudopotentials are used to capture the electron-ion interactions. The plane wave basis orbitals 5s$^2$5p$^6$6s$^2$ for Ba (Ba$\_$sv$\_$GW) and 2s$^2$2p$^4$ for O (O$\_$GW) are used as valence electrons. Plane wave cutoff energy and k-spacing are tested against the total energy to ensure the convergence. The plane wave cutoff energy was set to 600 eV and a k-spacing of 2$\pi$ $\times$ 0.024 $\AA^{-1}$  with $\Gamma$-centered Monkhorst-Pack scheme to compute structural and mechanical properties at ambient as well as at high pressure. To compute phonon dispersion curves, forces are calculated using finite displacement method for ambient and high pressure phases and then force constants are extracted and then post-processed by using phonopy package.~\cite{togo2015first} Pressure dependent Raman spectra is calculated using norm conserving pseudo potentials within linear response approach as implemented in CASTEP package\cite{Clark2005}. Crystal structures are visualized and analyzed using the VESTA software\cite{VESTA2008}. 
To include anharmonic effects, we also have calculated the finite temperature lattice dynamics for B2 phase of BaO using ab-initio molecular dynamics (AIMD) simulations and then post-processed using temperature dependent effective potential method (TDEP).\cite{TDEP_PRB_2011}

\section{Results and Discussion}
\subsection{Crystal structure and pressure-induced structural phase transitions}
\begin{figure}
\centering
\includegraphics[width=0.9\columnwidth]{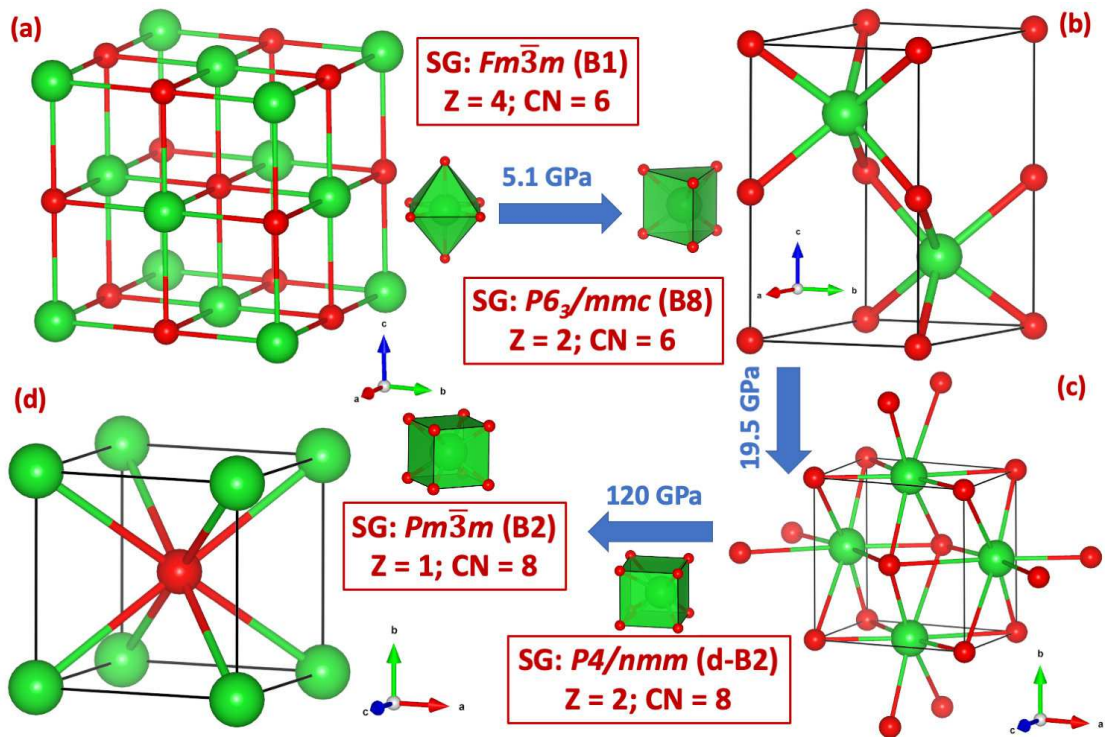}
\caption{Crystal structure of (a) ambient rocksalt NaCl-type (B1) and high pressure (b) NiAs-type (B8), (c) PH$_4$-type distorted CsCl-type (d-B2) and (d) CsCl-type (B2) phases of BaO. Where SG and CN denote space group and coordination number, respectively. Green and red color balls represent Ba and O atoms, respectively.}
\label{str}
\end{figure}

\begin{figure}
\centering
\includegraphics[width=0.95\columnwidth]{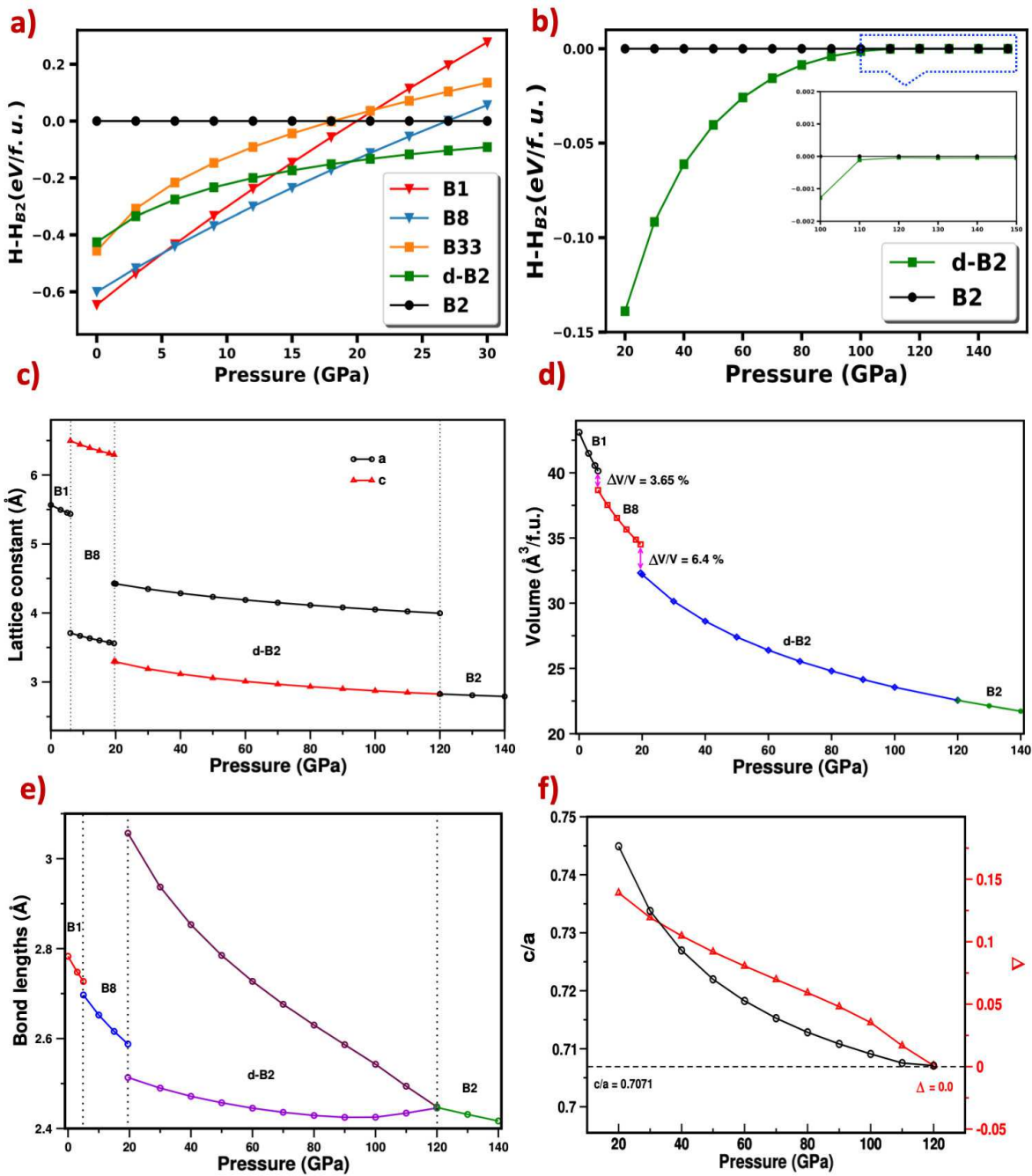}
\caption{Calculated (a,b) relative enthalpy difference of B1, B8, B33, d-B2 phases w.r.t B2 phase, (c) lattice constants, (d) volume, (e) bond lengths for B1, B8, DB2 and B2 phases as a function of pressure and (f) variation of c/a ratio and $\Delta$ with pressure for d-B2 phase.}
\label{lattice}
\end{figure}

As shown in Figure \ref{str}a, BaO crystallizes in the face centred cubic (SG: $Fm\bar{3}m$) rocksalt NaCl-type (B1) structure with 6-fold coordination and having Z = 4 formula units (f.u.) per unit cell at ambient conditions.\cite{Weir1986} High pressure X-ray diffraction measurements\cite{Liu1971,Liu1972,Weir1986}  revealed that BaO exhibits rich polymorphism under high pressure. BaO undergoes to a high pressure NiAs-type (B8) structure (SG: $P6_3/mmc$) from B1 phase at 9.2 GPa without changing the coordination number (CN) by having Z = 2 f.u. per primitive cell. The crystal structure of B8 phase is an hexagonal analog of the B1 structure. Later, the B8 phase transforms to PH$_4$-type distorted CsCl-type (d-B2) at 14.0 GPa,\cite{Liu1972} 18 GPa\cite{Liu1972} with pseudo 8-fold coordination (SG: $P4/nmm$) with Z = 2 f.u. per primitive cell, and then to a 8-fold coordinated CsCl-type (B2) structure (SG: $Pm\bar{3}m$) with Z = 1 f.u. per primitive cell upon further compression $i.e.,$ above 1 Mbar pressure (see Figure \ref{str} and Table \ref{table1}).\cite{Liu1971,Liu1972,Weir1986} The crystal structure of d-B2 phase is analogous to PbO-type structure without a lone-pair at Ba atom having two in-equivalent bond lengths as shown in Figure \ref{str}c. 
To get further insights on the observed structural phase transitions and their sequence, as a first step, we have tested with various DFT functionals such as LDA, PBE, PW91 and SCAN to reproduce the B1 to B8 transition pressure (see Figure S1). The transition pressure obtained using PBE, PW91 and SCAN functionals are consistent with each other while LDA poorly describe the phase stability between B1 and B8 phases. Later, we carried out static enthalpy calculations with PBE functional
for the polymorphic phases of BaO and plotted enthalpy difference with respect to B2 phase as a function of pressure as shown in Figure \ref{lattice}a. The obtained phase transition sequence is B1 $\rightarrow$ B8 $\rightarrow$ d-B2 $\rightarrow$ B2 and the corresponding transition pressures 5.1, 19.5, 120 GPa are in good agreement with the HP-XRD measurements\cite{Liu1972,Weir1986} as well as with previous first principles calculations\cite{Uludoan2001,Lukaevi2011-BaO}. Moreover, TlI-type (B33) phase is found to be meta-stable for CaO\cite{Catti2003} and we also have included this B33 phase in the phase diagram of BaO to verify it's relative phase stability w.r.t high pressure phases of BaO and then we found that the B33 phase is also meta-stable for BaO over the maximum studied pressure range (see Figure \ref{lattice}a). As presented in Figure \ref{lattice}b, the enthalphy difference between d-B2 and B2 phases is continuously decreasing with increasing pressure and they become iso-enthalpic at around 120 GPa. As shown in Figure \ref{lattice}c, the calculated lattice constants for the ambient and high pressure phases are monotonically decreasing with pressure. The lattice constant 'c' of d-B2 phase becomes lattice constant 'a' for B2 phase (c$_{d-B2}$ = a$_{B2}$). The detailed transition mechanism for d-B2 to B2 transition is explained through elastic constants in the following section. As shown in Figure \ref{lattice}d, the calculated volume decreases monotonically with pressure. We observe a volume collapse during the phase transition from B1 $\rightarrow$ B8 is $\sim$ 3.65 $\%$ at 5.1 GPa and it is $\sim$ 6.4 $\%$ at 19.5 GPa for the transition B8 $\rightarrow$ d-B2. The obtained volume reduction is consistent with high pressure X-diffraction measurements and previous first principles calculations as presented in Table \ref{table2}. This clearly demonstrates that B1 to B8 $\&$ B8 to d-B2 transitions are first order in nature, while d-B2 to B2 transition is a second order or weak first order in nature due to continuous change in volume during the phase transition (see Figure \ref{lattice}d). We then calculated the equilibrium bulk modulus (B$_0$) and its pressure derivative (B$_0$') by fitting the obtained pressure-volume (P-V) data to 3$^{rd}$ order Birch-Murnaghan equation of state,\cite{EOS_1947} B$_0$ measures the ability of a material to resist changes in volume under hydrostatic expansion or compression. The calculated B$_0$ value is found to be 71.4 GPa for the ambient B1 phase, which is in good agreement with previous first principles calculations.\cite{MO-Elastic-CINTHIA201523} The obtained relatively low B$_0$ values  68.8 GPa for B8 and 51.8 for d-B2 intermediate phases suggest the soft nature of these phases over ambient B1 (71.4 GPa) and a high B$_0$ value 92.4 GPa for high pressure B2 phase disclose that B2 phase is harder than B1 phase (see Table \ref{table3}). 

In addition, the calculated in-equivalent bond lengths are plotted as a function of pressure and the same are presented in Figure \ref{lattice}e. The in-equivalent bond lengths of d-B2 phase are found to converge to became a single bond length for B2 phase above 120 GPa. In addition, in the d-B2 phase, the Ba atoms are located at (0.5, 0.0, 0.5+$\Delta$) and (0.0, 0.5, 0.5-$\Delta$), where $\Delta$ is the atomic position parameter, $\Delta$ in combination axial ratio (c/a) provides a useful information on d-B2 $\rightarrow$ B2 transition. If $\Delta$ = 0 $\&$ c/a = 1/$\sqrt2$ = 0.7071, then the structure is of undistorted B2 phase.\cite{Weir1986} To confirm this, we have calculated c/a ratio and $\Delta$ as a function of pressure for d-B2 phase. As shown in Figure \ref{lattice}f, both c/a ratio and $\Delta$ are decreasing with pressure and then approaches to  $\Delta$ = 0 $\&$ c/a = 0.7071 around 120 GPa. 
The continuous volume change at the transition pressure, variation in the fractional coordinate of Ba atom along c-direction (see Figure S2), c/a = 0.7071 and $\Delta$ = 0 values clearly demonstrate that d-B2 to B2 transition to be a displacive transition under pressure. Displacive nature of $z_{Ba}$ with pressure suggests phonon softening in d-B2 phase as function of pressure.
Further to explore the microscopic origin of the continuous phase transition in BaO, pressure dependent elastic constants, lattice dynamics and Raman spectra are calculated and analyzed them in detail in the following sections.

\subsection{Elastic constants under high pressure}
To get deeper insight on the structural phase transition(s) and verify mechanical stability of the BaO polymorphs, we have calculated second order elastic constants (C$_{ij}$) as a function of pressure for all the four B1, B8, d-B2 and B2 phases. The cubic ($Fm\bar{3}m$, $Pm\bar{3}m$), hexagonal ($P6_3/mmc$) and tetragonal ($P4/nmm$) crystal symmetries possess 3, 5 and 6 independent elastic constants. 

\begin{figure}
\centering
\includegraphics[width=6.5in,height=3.0in]{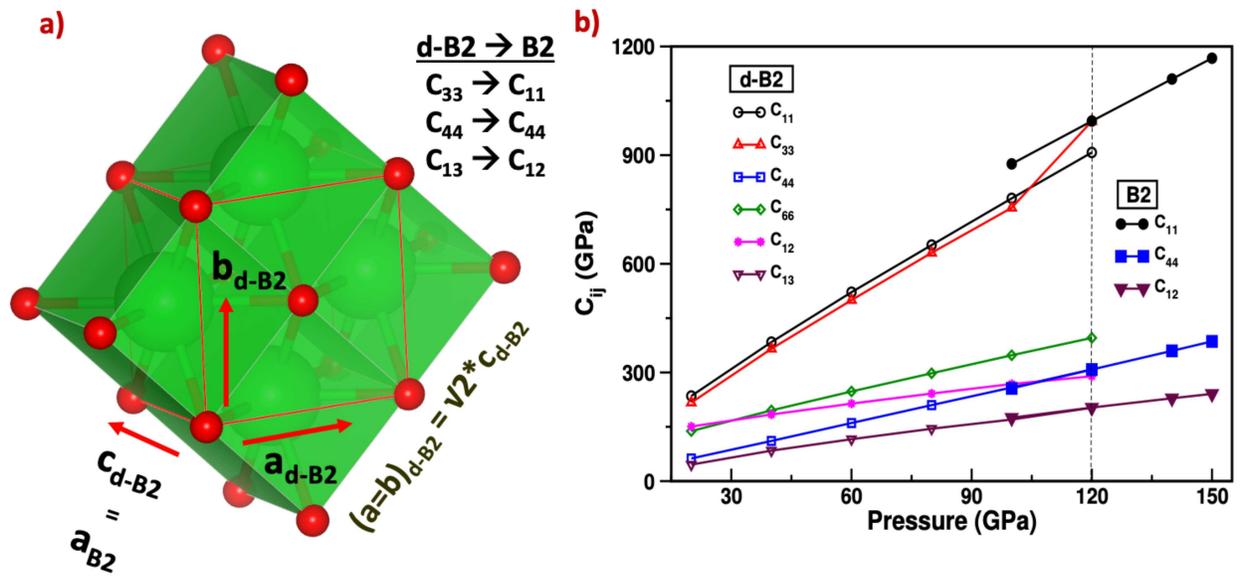}
\caption{(a) Lattice transformation from d-B2 to B2 and (b) calculated elastic constants of d-B2 and B2 phases of BaO as a function of pressure.}
\label{Elastic}
\end{figure}

When a finite hydrostatic pressure is applied to a crystal system, the  Born stability criteria\cite{Mouhat2014} at ambient conditions must be revised with modified elastic constants at a given hydrostatic pressure conditions. According to Sinko and Smirnov the modified elastic constants at a given pressure (P) are $\widetilde{C}_{ii}$ = $C_{ii}$ - P, (for $i$ = 1-6) and  $\widetilde{C}_{1j}$ = $C_{1j}$ + P (for $j$ = 2-3) for cubic and hexagonal/tetragonal systems. Therefore, the modified Born stability criteria under hydrostatic pressure for cubic (equation 1) and hexagonal/tetragonal (equation 2) are given as follows:
\begin{equation}
C_{11} - C_{12} - 2P  >  0, \hspace{0.2in}  C_{11} + 2C_{12} + P  >  0, \hspace{0.2in} C_{44} - P  >  0
\end{equation}
\begin{equation}
\begin{split}
C_{11} - C_{12} - 2P  >  0, \\ P^2 + P (C_{11} + C_{12} + 4C_{13}) + 2C^2_{13} - C_{33}(C_{11} + C_{12}) > 0, \\  C_{44} - P  >  0, \hspace{0.2in} C_{66} - P > 0
\end{split}
\end{equation}
The calculated pressure dependent C$_{ij}$ for B1, B8, d-B2 and B2 phases satisfy the necessary and sufficient conditions for the revised Born stability criteria under hydrostatic pressure indicating that the polymoprhs are mechanically stable at the corresponding pressures. We shed more light on understanding lattice transformation from d-B2 to B2 phase through the computed pressure dependent elastic constants of d-B2 and B2 phases. The relationships between lattice constants of d-B2 and B2 phases at the phase transition are given as follows: (a=b)$_{d-B2}$ = $\sqrt2c_{d-B2}$ and c$_{d-B2}$ = a$_{B2}$ (see Figure \ref{Elastic}a), thus, results in merging of the following elastic constants of d-B2 to B2 phase under pressure are as depicted in Figure \ref{Elastic} as given below: C$^{d-B2}_{33}$ $\rightarrow$ C$^{B2}_{11}$, C$^{d-B2}_{44}$ $\rightarrow$ C$^{B2}_{44}$ and C$^{d-B2}_{13}$ $\rightarrow$ C$^{B2}_{12}$, which corroborates with pressure dependent structural and bond parameters at the d-B2 $\rightarrow$ B2 transition.


\section{Lattice dynamics and Raman spectra under high pressure}
\begin{figure}
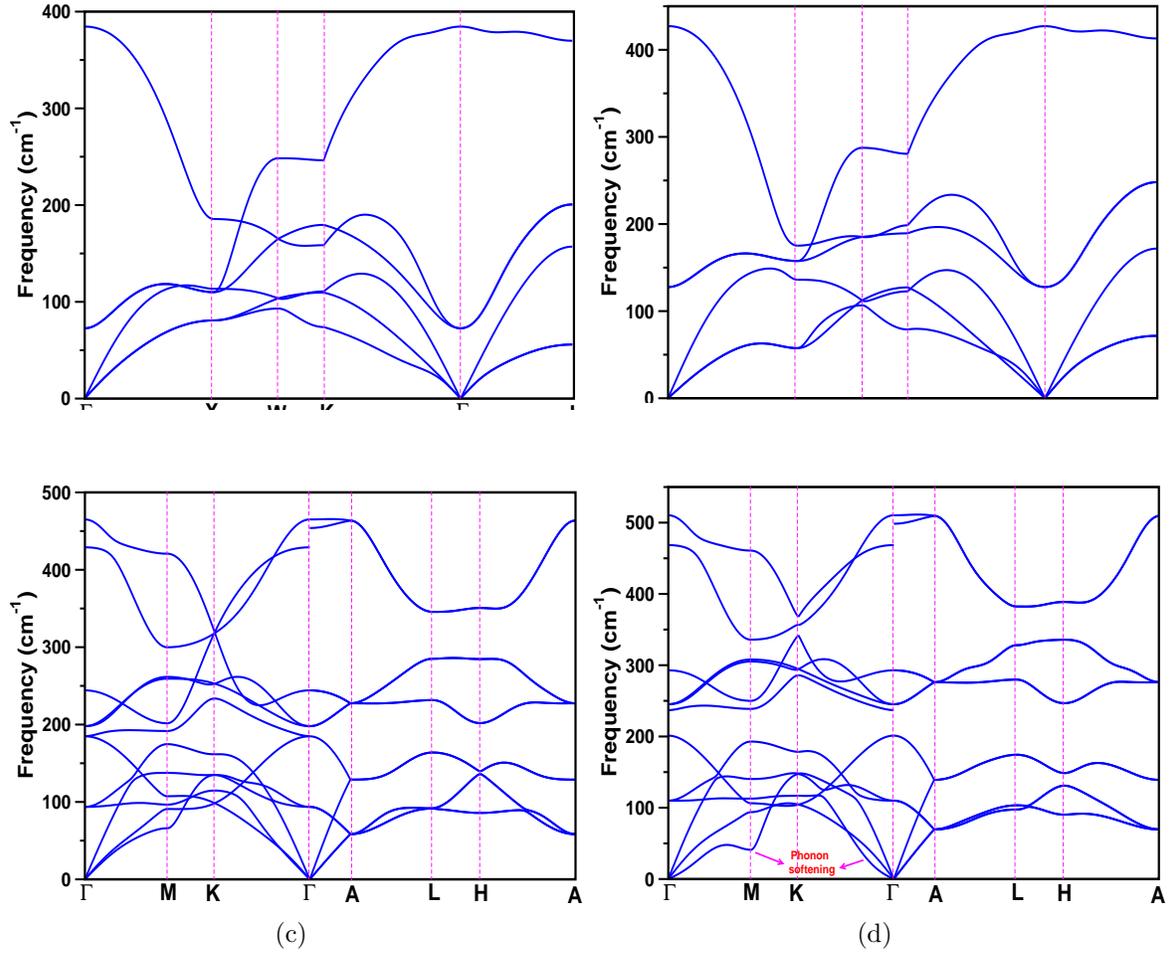

\centering
\subfigure[]{\includegraphics[width=3.0in,height=2.2in]{Figures/BaO-225-0GPa.eps}} 
\subfigure[]{\includegraphics[width=3.0in,height=2.2in]{Figures/BaO-225-9GPa.eps}} 
\subfigure[]{\includegraphics[width=3.0in,height=2.2in]{Figures/BaO-194-9GPa.eps}} 
\subfigure[]{\includegraphics[width=3.0in,height=2.2in]{Figures/BaO-194-20GPa.eps}} 
\caption{Calculated phonon dispersion curves of (a, b) B1 phase at (a) 0, (b) 9 GPa  and for (c, d) B8 phase at (c) 9 GPa, (d) 20 GPa. Phonon softening is observed along X-direction for B1 phase and for B8 phase at high pressure (20 GPa) along M and K-$\Gamma$ directions, which destabilize the B1 and B8 phases upon further compression.}
\label{fig:PD-225}
\end{figure}

\begin{figure}
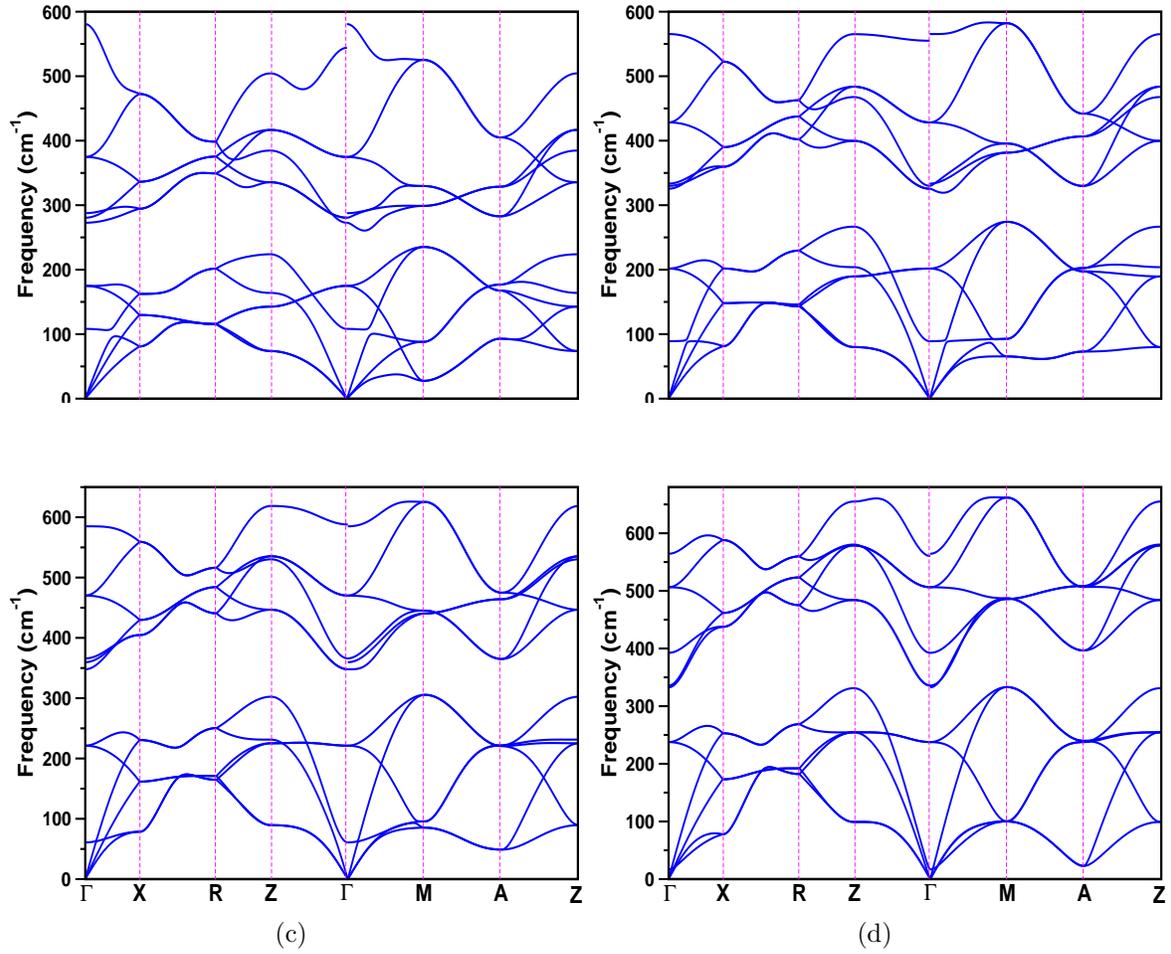

\centering
\subfigure[]{\includegraphics[width=3.0in,height=2.2in]{Figures/BaO-129-30GPa.eps}} 
\subfigure[]{\includegraphics[width=3.0in,height=2.2in]{Figures/BaO-129-60GPa.eps}} 
\subfigure[]{\includegraphics[width=3.0in,height=2.2in]{Figures/BaO-129-90GPa.eps}} 
\subfigure[]{\includegraphics[width=3.0in,height=2.2in]{Figures/BaO-129-120GPa.eps}} 
\caption{Calculated phonon dispersion curves for d-B2 phase at (a) 30, (b) 60, (c) 90 and (d) 120 GPa. Optical phonon softening along Z-$\Gamma$ direction and acoustic phonon softening along A-direction are observed with pressure and it is predominant above 90 GPa.}
\label{fig:PD-129}
\end{figure}

\begin{figure}
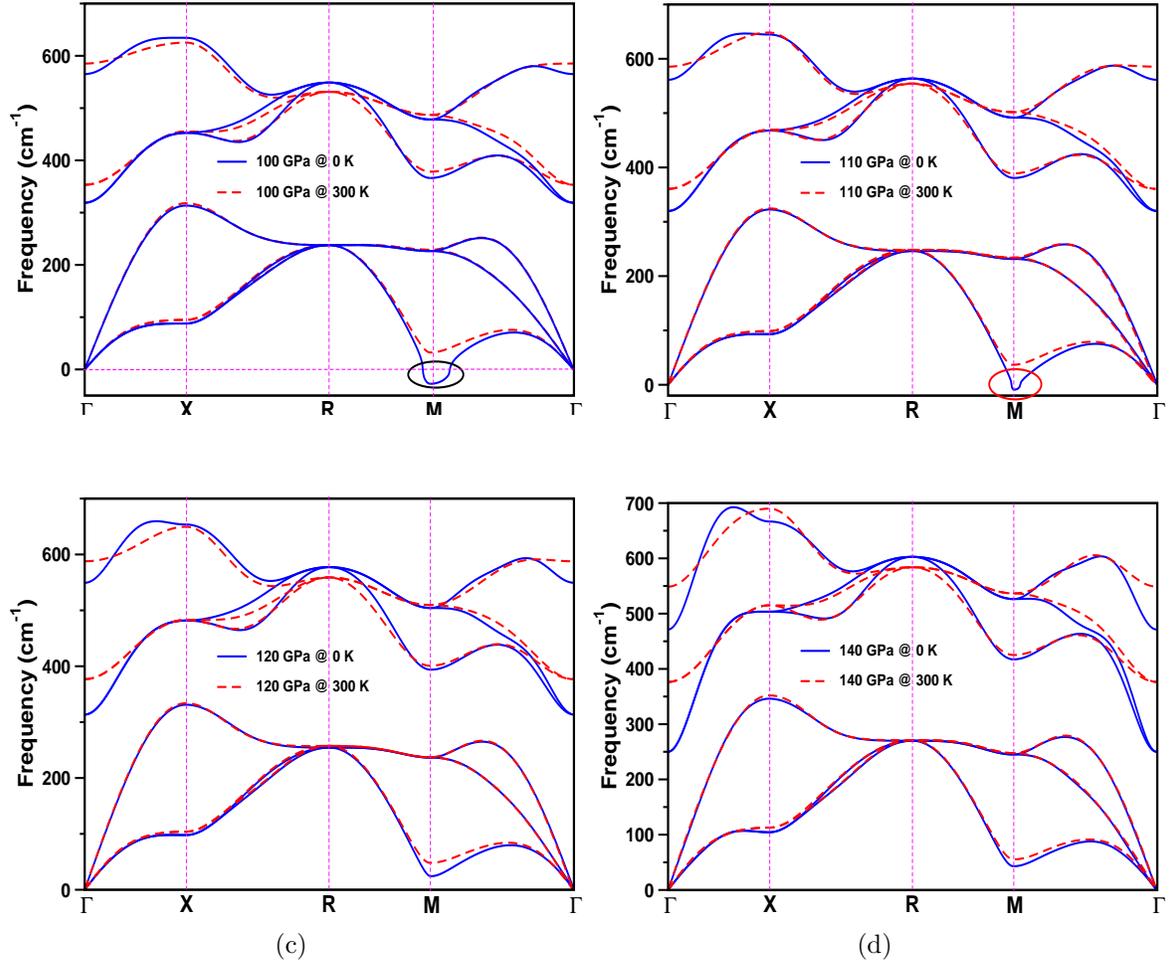

\centering
\subfigure[]{\includegraphics[width=3.0in,height=2.2in]{Figures/BaO-221-100GPa-New.eps}}  
\subfigure[]{\includegraphics[width=3.0in,height=2.2in]{Figures/BaO-221-110GPa-New.eps}}  \\
\subfigure[]{\includegraphics[width=3.0in,height=2.2in]{Figures/BaO-221-120GPa-New.eps}}  
\subfigure[]{\includegraphics[width=3.0in,height=2.2in]{Figures/BaO-221-140GPa-New.eps}}    
\caption{Calculated phonon dispersion curves for B2 phase at high pressures (a) 100 GPa, (b) 110 GPa (c) 120 GPa and (d) 140 GPa at 0K as well as at 300 K. B2 phase is dynamically stable including anharmonic effects at 100 and 110 GPa at 300 K but it is dynamically unstable within the harmonic approximation at 0 K.}
\label{fig:PD-221}
\end{figure}

To compliment the structural stability obtained from total energy calculations and verify dynamical stability of the ambient and high pressure phases of BaO, we calculated phonon dispersion curves including LO-TO splitting at 0 and 9 GPa for B1 phase (see Figure \ref{fig:PD-225}a $\&$ b) at 9 GPa and 20 GPa for B8 phase (see Figure \ref{fig:PD-225}c $\&$ d), at 30, 60, 90 and 120 GPa for d-B2 phase (see Figure \ref{fig:PD-129}) and finally at 100, 110, 120 and 140 GPa for B2 phase (see Figure \ref{fig:PD-221}). As shown in Figure \ref{fig:PD-225}a $\&$ b, B1 and B8 phases are dynamically stable at 0, 9 GPa and 9, 20 GPa respectively. The phonon dispersion curves of B1 phase show large LO-TO splitting ($\sim$ 312.3 cm$^{-1}$) along $\Gamma$-direction with two degenerate TO and one LO modes and this large LO-TO splitting causes a fall of low lying optical (LLO) phonon modes into acoustic mode region, which can increase overlap between acoustic and LLO phonon modes there by enhancing the scattering phase space and eventually this leads to lower the lattice thermal conductivity in B1 phase at ambient conditions.\cite{Roshan2021} However, the LO-TO splitting decreases under pressure (see Figure \ref{fig:PD-225}a $\&$ b) for B1 phase. Moreover, the optical phonon modes are hardening with pressure while phonon softening is observed for acoustic phonon modes along X-direction in B1 phase and along M and K-$\Gamma$ directions in B8 phase. Phonon softening is observed along X-direction for B1 phase previously,\cite{Lukaevi2011-BaO} which actually drives the transition from B1 $\rightarrow$ B8 phase in BaO under pressure. In addition, the B1 and B8 phases are dynamically stable above the transition pressures, which are in consistent with the first order nature of the structural phase transitions, B1 $\rightarrow$ B8  $\&$  B8 $\rightarrow$ d-B2. The d-B2 phase is found to be dynamically stable over a wide pressure range of $\sim$ 20-120 GPa (see Figure \ref{fig:PD-129}) for BaO in contrast to the light alkaline-earth metal oxides, MgO, CaO and SrO. The calculated phonon dispersion curves show a significant phonon softening at 120 GPa along Z-$\Gamma$ and 'A' high symmetry directions as depicted in Figure \ref{fig:PD-129}d. On the other hand, the calculated phonon dispersion curves for B2 phase at 100 and 110 GPa show soft phonon mode along M high symmetry point, which clearly indicates that B2 is unstable below 110 GPa pressure. In addition, we could also observe a significant phonon softening of TO and LO modes of B2 phase at 140 GPa (see Figure \ref{fig:PD-221}d) compared to 120 GPa (see Figure \ref{fig:PD-221}b), which indicates that B2 phase becomes dynamically unstable and it may undergo a structural phase transition upon further compression above 150 GPa. 

\begin{figure}
\centering
\includegraphics[width=0.8\columnwidth]{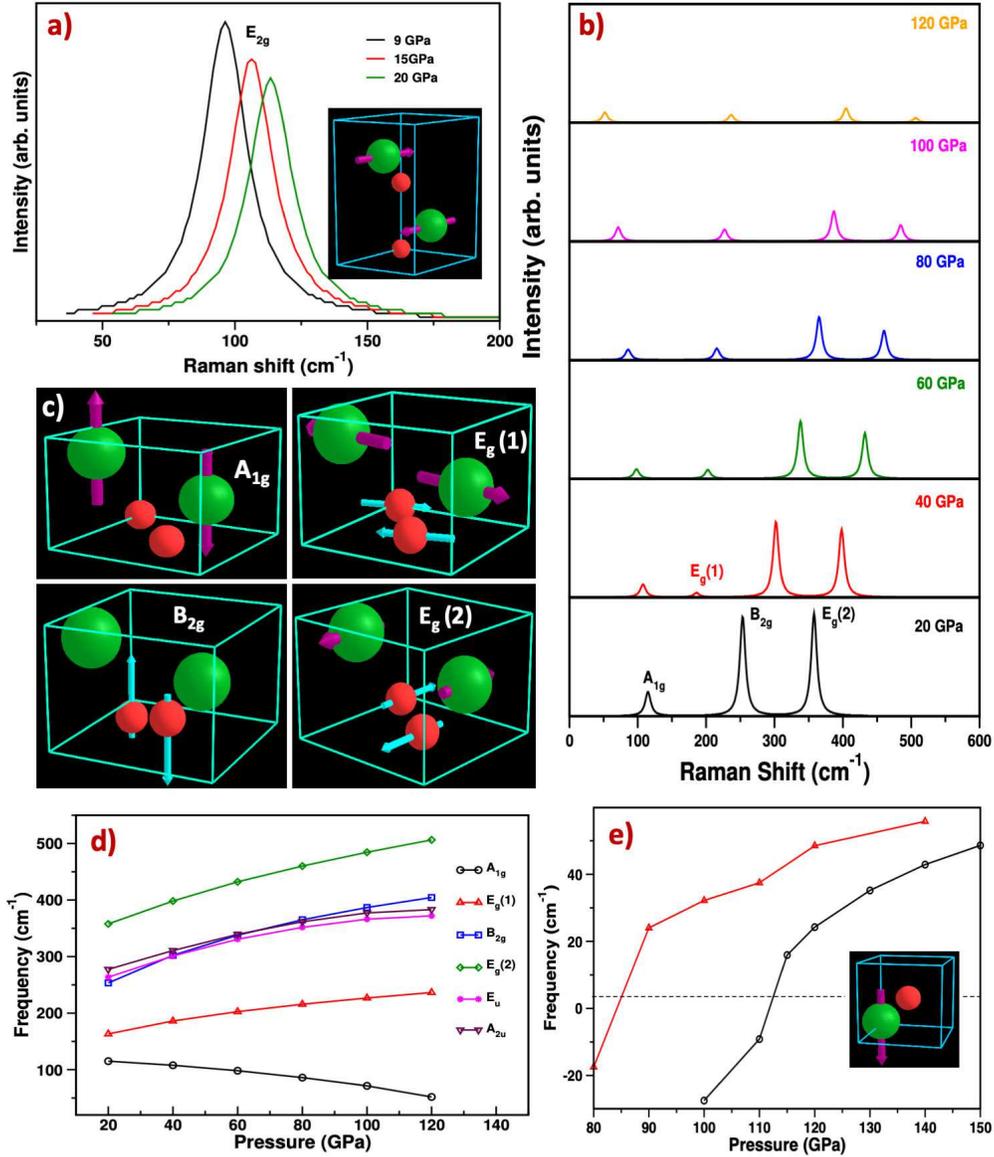}
\caption{Calculated Raman spectra of (a) B8 and (b) d-B2 phase as a function of pressure. Eigen vectors of Raman active modes are presented for (a) B8 and (c) d-B2 phases of BaO. Pressure evolution of (d) Raman and IR active modes of d-B2 phase and (e) Transverse acoustic mode of B2 phase along M-high symmetry direction without and with inclusion of anharmonic effects.}
\label{fig:Raman}
\end{figure}

The cubic ambient B1 and high pressure B2 phases of BaO are Raman inactive, and they have only one IR active (T$_{1u}$) mode. However, the intermediate high pressure B8 and d-B2 phases are Raman active, and are thoroughly analyzed (see Figure \ref{fig:Raman}). 
This result is in very good agreement with the Raman spectroscopic measurements,\cite{Efthimiopoulos2010} which show that BaO is Raman inactive until 10 GPa (B1 phase) and above 11 GPa it becomes Raman active (B8), where one Raman active E$_{2g}$ mode is observed and this Raman mode shows blue-shift with increasing pressure. Later, significant changes are observed in the Raman spectra at around 14 GPa and it indicates a transition from B8 to to d-B2 phase. The similar trends are observed in the present work with deviations in the transition pressures which might be due to Raman measurements were carried out at 300 K, while the calculations are performed at 0 K.
Since, B8 and d-B2 phases crystallize in $P6_3/mmc$ and $P4/nmm$ space groups, respectively and they consist of four atoms per primitive cell, the symmetry decomposition of group theory predicts 12 vibrational modes along $\Gamma$ for both of these phases, which are given as follows: 
\begin{center}
$\Gamma_{12}^{B8}$ = 2E$_{2g}$ $\oplus$ B$_{2g}$ $\oplus$ 4E$_{1u}$ $\oplus$ 2E$_{2u}$ $\oplus$ 2A$_{2u}$ $\oplus$ B$_{1u}$ 
\end{center}
Among the 12 vibrational modes of B8 phase, only E$_{2g}$ is Raman active mode which is consistent with high pressure Raman measurements,\cite{Efthimiopoulos2010} E$_{1u}$, E$_{2u}$, A$_{2u}$ and B$_{1u}$ are IR active modes but B$_{2g}$ is neither Raman nor IR active (silent) mode; where E$_{1u}$, E$_{2u}$ and E$_{2g}$ are doubly degenerate modes. Variation of E$_{2g}$ mode as a function of pressure is shown in Figure \ref{fig:Raman}a, which is mainly due to in-plane vibrational motion of Ba atom and its intensity decreases with pressure. 
\begin{center}
$\Gamma_{12}^{d-B2}$ = 4E$_g$ $\oplus$ B$_{2g}$ $\oplus$ A$_{1g}$ $\oplus$ 4E$_{u}$ $\oplus$ 2A$_{2u}$  
\end{center}
On the other hand, for Litharge PbO-type d-B2 phase, the zone-center Raman active modes  B$_{2g}$, A$_{1g}$ and  E$_g$ are polarized along c- and a/b-axis, respectively.
E$_{u}$ and A$_{2u}$ are IR active modes; where E$_{u}$ and E$_{g}$ modes are doubly degenerate. Pressure dependence of vibrational frequency modes are presented in Figure \ref{fig:Raman}d. For each vibrational mode, frequency increases with pressure except for A$_{1g}$ mode. When the pressure is applied, the interatomic distance becomes smaller and hence interaction between neighboring atoms becomes stronger. Therefore, usually, the vibrational modes increase with pressure as shown in Figure \ref{fig:Raman}b $\&$ d except for A$_{1g}$ mode. As E$_{g}$ mode is in plane vibrational mode, the vibrational frequency/wavenumber increases as the interaction of atoms becomes stronger. As shown in Figure \ref{lattice}f; c/a ratio decreases with increasing pressure. Due to the out-of-plane vibrations of heavier Ba atoms in A$_{1g}$ mode (Figure \ref{fig:Raman}c), the decrease in the wavenumber of Raman-shift is observed with pressure. The red-shift of the A$_{1g}$ mode with pressure shows a negative pressure coefficient and it might be responsible for introducing dynamical instability at higher pressure in the d-B2 phase. 
In contrast, this mode was tentatively assigned as E$_g$ mode due to in-plane vibration of Ba atoms in the HP-Raman study.\cite{Efthimiopoulos2010}. However, we thoroughly analyzed and assigned it as A$_{1g}$ mode rather than E$_g$ mode, which actually shows phonon softening with pressure which is associated with displacive nature of Ba atoms along c-axis.
The Intensity of four Raman active modes (E$_g$, B$_{2g}$ and A$_{1g}$) are gradually diminishing with pressure (Figure \ref{fig:Raman}b) for d-B2 phase, this strongly suggests that high pressure phase might be Raman inactive. Finally, the high pressure B2 phase can be stabilized above 113 GPa (see Figure \ref{fig:Raman}e) which is consistent with the observations from the HP-XRD measurements.\cite{Weir1986}.

As shown in Figure S3 $\&$ $\ref{fig:PD-221}$, the calculated room temperature phonon dispersion curves in combination with AIMD and TDEP disclose that B2 phase is dynamically unstable at 80 GPa and 300 K whereas it becomes dynamically stable at 100 GPa and 300 K, while harmonic lattice dynamical calculations show that B2 phase is dynamically unstable below $\sim$ 113 GPa. This clearly shows the importance of anharmonicity to determine the stability of B2 phase above 85 GPa and this is in good accord with the HP-XRD diffraction measurements.\cite{Weir1986}


\section{Conclusions}
To summarize, we have systematically investigated the polymorphism of BaO under hydrostatic compression using first principles calculations based on density functional theory. We found that BaO undergoes series of pressure-induced structural phase transitions from B1 $\rightarrow$  B8 $\rightarrow$  d-B2 $\rightarrow$  B2 and the corresponding transition pressures are 5.1 , 19.5, 120 GPa respectively. The obtained structural phase transition sequence and transition pressures are consistent with HP-XRD measurements and previous first principles calculations. TlI-type (B33) phase is found to be metastable for BaO over the studied pressure range. The d-B2 phase exhibits stability over a wide pressure range $\sim$ 19.5-113 GPa in contrast to light alkaline-earth metal oxides. The transitions from B1 $\rightarrow$ B8 and B8 $\rightarrow$ d-B2 are first order in nature with a volume collapse of $\sim$ 3.65 and 6.4 $\%$, respectively while d-B2$\rightarrow$ B2 is a second order or weak first order phase transition with displacive nature. Mechanical stability has been established for ambient and high pressure phases of BaO through computed elastic constants. The pressure dependent elastic constants provide an insight on lattice transformation from d-B2 to B2 phase. The calculated phonon dispersion curves indicate the dynamical stability of BaO polymorphs and a significant phonon softening is observed for B8 and d-B2 phases at high pressures and also a soft phonon mode (along M-direction) in B2 phase explained its dynamical instability below 110 GPa. Finally, gradual diminishing of Raman active modes of d-B2 phase under pressure suggests that high pressure phase could be a Raman inactive. Overall, the present study provides an in depth understanding of structural phase transitions in BaO under high pressure and further exploration of high pressure behavior of light alkaline-earth oxides (MgO, CaO and SrO) have important implications on composition of the Earth's lower mantle.

\section{Acknowledgments}
LV and NYK contributed equally to this manuscript. NYK would like to thank Science and Engineering Research Board and Indo-US Scientific Technology Forum for providing financial support through SERB Indo-US postdoctoral fellowship and Institute for Advanced Computational Science, Stony Brook University for providing computational resources (Seawulf cluster). SCRR would like thank RGUKT Basar for providing computational facilities.  

\bibliography{references.bib}

\providecommand{\latin}[1]{#1}
\makeatletter
\providecommand{\doi}
  {\begingroup\let\do\@makeother\dospecials
  \catcode`\{=1 \catcode`\}=2 \doi@aux}
\providecommand{\doi@aux}[1]{\endgroup\texttt{#1}}
\makeatother
\providecommand*\mcitethebibliography{\thebibliography}
\csname @ifundefined\endcsname{endmcitethebibliography}
  {\let\endmcitethebibliography\endthebibliography}{}
\begin{mcitethebibliography}{77}
\providecommand*\natexlab[1]{#1}
\providecommand*\mciteSetBstSublistMode[1]{}
\providecommand*\mciteSetBstMaxWidthForm[2]{}
\providecommand*\mciteBstWouldAddEndPuncttrue
  {\def\EndOfBibitem{\unskip.}}
\providecommand*\mciteBstWouldAddEndPunctfalse
  {\let\EndOfBibitem\relax}
\providecommand*\mciteSetBstMidEndSepPunct[3]{}
\providecommand*\mciteSetBstSublistLabelBeginEnd[3]{}
\providecommand*\EndOfBibitem{}
\mciteSetBstSublistMode{f}
\mciteSetBstMaxWidthForm{subitem}{(\alph{mcitesubitemcount})}
\mciteSetBstSublistLabelBeginEnd
  {\mcitemaxwidthsubitemform\space}
  {\relax}
  {\relax}

\bibitem[Cho \latin{et~al.}(1999)Cho, Kim, Lee, Yeom, Kim, and Park]{Cho1999}
Cho,~J.; Kim,~R.; Lee,~K.-W.; Yeom,~G.-Y.; Kim,~J.-Y.; Park,~J.-W. Effect of
  CaO addition on the firing voltage of MgO films in AC plasma display panels.
  \emph{Thin Solid Films} \textbf{1999}, \emph{350}, 173--177\relax
\mciteBstWouldAddEndPuncttrue
\mciteSetBstMidEndSepPunct{\mcitedefaultmidpunct}
{\mcitedefaultendpunct}{\mcitedefaultseppunct}\relax
\EndOfBibitem
\bibitem[Speier and Szot(1997)Speier, and Szot]{Speier1997}
Speier,~W.; Szot,~K. Physics and chemistry at oxide surfaces. {ByClaudine}
  Noguera, Cambridge university press, Cambridge 1996, xv, 223 pp., hardcover,
  {\textsterling}40.00, {ISBN} 0-52147214-8. \emph{Advanced Materials}
  \textbf{1997}, \emph{9}, 1192--1193\relax
\mciteBstWouldAddEndPuncttrue
\mciteSetBstMidEndSepPunct{\mcitedefaultmidpunct}
{\mcitedefaultendpunct}{\mcitedefaultseppunct}\relax
\EndOfBibitem
\bibitem[Duan and Sorescu(2010)Duan, and Sorescu]{Duan2010}
Duan,~Y.; Sorescu,~D.~C. CO$_2$capture properties of alkaline earth metal
  oxides and hydroxides: A combined density functional theory and lattice
  phonon dynamics study. \emph{The Journal of Chemical Physics} \textbf{2010},
  \emph{133}, 074508\relax
\mciteBstWouldAddEndPuncttrue
\mciteSetBstMidEndSepPunct{\mcitedefaultmidpunct}
{\mcitedefaultendpunct}{\mcitedefaultseppunct}\relax
\EndOfBibitem
\bibitem[Medeiros \latin{et~al.}(2007)Medeiros, Albuquerque, Maia, de~Sousa,
  Caetano, and Freire]{Medeiros2007}
Medeiros,~S.~K.; Albuquerque,~E.~L.; Maia,~F.~F.; de~Sousa,~J.~S.; Caetano,~E.
  W.~S.; Freire,~V.~N. {CaO} first-principles electronic properties and {MOS}
  device simulation. \emph{Journal of Physics D: Applied Physics}
  \textbf{2007}, \emph{40}, 1655--1658\relax
\mciteBstWouldAddEndPuncttrue
\mciteSetBstMidEndSepPunct{\mcitedefaultmidpunct}
{\mcitedefaultendpunct}{\mcitedefaultseppunct}\relax
\EndOfBibitem
\bibitem[Zannotti \latin{et~al.}(2015)Zannotti, Wood, Summers, Stevens, Hall,
  Snape, Giovanetti, and Gibson]{Zannotti2015}
Zannotti,~M.; Wood,~C.~J.; Summers,~G.~H.; Stevens,~L.~A.; Hall,~M.~R.;
  Snape,~C.~E.; Giovanetti,~R.; Gibson,~E.~A. Ni Mg Mixed Metal Oxides for
  p-Type Dye-Sensitized Solar Cells. \emph{{ACS} Applied Materials {\&}
  Interfaces} \textbf{2015}, \emph{7}, 24556--24565\relax
\mciteBstWouldAddEndPuncttrue
\mciteSetBstMidEndSepPunct{\mcitedefaultmidpunct}
{\mcitedefaultendpunct}{\mcitedefaultseppunct}\relax
\EndOfBibitem
\bibitem[Prinetto \latin{et~al.}(2003)Prinetto, Ghiotti, Nova, Castoldi,
  Lietti, Tronconi, and Forzatti]{app-1-Prinetto2003}
Prinetto,~F.; Ghiotti,~G.; Nova,~I.; Castoldi,~L.; Lietti,~L.; Tronconi,~E.;
  Forzatti,~P. In situ {FT}-{IR} and reactivity study of {NOxstorage} over
  Pt{\textendash}Ba/Al$_2$O$_3$catalysts. \emph{Phys. Chem. Chem. Phys.}
  \textbf{2003}, \emph{5}, 4428--4434\relax
\mciteBstWouldAddEndPuncttrue
\mciteSetBstMidEndSepPunct{\mcitedefaultmidpunct}
{\mcitedefaultendpunct}{\mcitedefaultseppunct}\relax
\EndOfBibitem
\bibitem[Choudhary \latin{et~al.}(1998)Choudhary, Rajput, and
  Mamman]{Choudhary1998}
Choudhary,~V.; Rajput,~A.; Mamman,~A. {NiO}-Alkaline Earth Oxide Catalysts for
  Oxidative Methane-to-Syngas Conversion: Influence of Alkaline Earth Oxide on
  the Surface Properties and Temperature-Programmed Reduction/Reaction by H2and
  Methane. \emph{Journal of Catalysis} \textbf{1998}, \emph{178},
  576--585\relax
\mciteBstWouldAddEndPuncttrue
\mciteSetBstMidEndSepPunct{\mcitedefaultmidpunct}
{\mcitedefaultendpunct}{\mcitedefaultseppunct}\relax
\EndOfBibitem
\bibitem[Au \latin{et~al.}(1998)Au, Chen, and Ng]{Au1998}
Au,~C.; Chen,~K.; Ng,~C. The modification of Gd$_2$O$_3$ with {BaO} for the
  oxidative coupling of methane reactions. \emph{Applied Catalysis A: General}
  \textbf{1998}, \emph{170}, 81--92\relax
\mciteBstWouldAddEndPuncttrue
\mciteSetBstMidEndSepPunct{\mcitedefaultmidpunct}
{\mcitedefaultendpunct}{\mcitedefaultseppunct}\relax
\EndOfBibitem
\bibitem[Mishra \latin{et~al.}(2004)Mishra, Garner, and
  Schmidt]{app2-Mishra2004}
Mishra,~K.~C.; Garner,~R.; Schmidt,~P.~C. Model of work function of tungsten
  cathodes with barium oxide coating. \emph{Journal of Applied Physics}
  \textbf{2004}, \emph{95}, 3069--3074\relax
\mciteBstWouldAddEndPuncttrue
\mciteSetBstMidEndSepPunct{\mcitedefaultmidpunct}
{\mcitedefaultendpunct}{\mcitedefaultseppunct}\relax
\EndOfBibitem
\bibitem[McKee \latin{et~al.}(2001)McKee, Walker, and Chisholm]{McKee2001}
McKee,~R.~A.; Walker,~F.~J.; Chisholm,~M.~F. Physical Structure and Inversion
  Charge at a Semiconductor Interface with a Crystalline Oxide. \emph{Science}
  \textbf{2001}, \emph{293}, 468--471\relax
\mciteBstWouldAddEndPuncttrue
\mciteSetBstMidEndSepPunct{\mcitedefaultmidpunct}
{\mcitedefaultendpunct}{\mcitedefaultseppunct}\relax
\EndOfBibitem
\bibitem[Chen \latin{et~al.}(1998)Chen, Liebermann, and Weidner]{Chen1998}
Chen,~G.; Liebermann,~R.~C.; Weidner,~D.~J. Elasticity of Single-Crystal {MgO}
  to 8 Gigapascals and 1600 Kelvin. \emph{Science} \textbf{1998}, \emph{280},
  1913--1916\relax
\mciteBstWouldAddEndPuncttrue
\mciteSetBstMidEndSepPunct{\mcitedefaultmidpunct}
{\mcitedefaultendpunct}{\mcitedefaultseppunct}\relax
\EndOfBibitem
\bibitem[Margrave(1979)]{mantle-Margrave1979}
Margrave,~J. High pressure research, applications in geophysics. \emph{Physics
  of the Earth and Planetary Interiors} \textbf{1979}, \emph{20}, 74\relax
\mciteBstWouldAddEndPuncttrue
\mciteSetBstMidEndSepPunct{\mcitedefaultmidpunct}
{\mcitedefaultendpunct}{\mcitedefaultseppunct}\relax
\EndOfBibitem
\bibitem[Bovolo(2005)]{Bovolo2005}
Bovolo,~C.~I. The physical and chemical composition of the lower mantle.
  \emph{Philosophical Transactions of the Royal Society A: Mathematical,
  Physical and Engineering Sciences} \textbf{2005}, \emph{363},
  2811--2836\relax
\mciteBstWouldAddEndPuncttrue
\mciteSetBstMidEndSepPunct{\mcitedefaultmidpunct}
{\mcitedefaultendpunct}{\mcitedefaultseppunct}\relax
\EndOfBibitem
\bibitem[Anderson(1989)]{anderson1989theory}
Anderson,~D.~L. \emph{Theory of the Earth}; Blackwell scientific publications,
  1989\relax
\mciteBstWouldAddEndPuncttrue
\mciteSetBstMidEndSepPunct{\mcitedefaultmidpunct}
{\mcitedefaultendpunct}{\mcitedefaultseppunct}\relax
\EndOfBibitem
\bibitem[McDonough and s.~Sun(1995)McDonough, and s.~Sun]{McDonough1995}
McDonough,~W.; s.~Sun,~S. The composition of the Earth. \emph{Chemical Geology}
  \textbf{1995}, \emph{120}, 223--253\relax
\mciteBstWouldAddEndPuncttrue
\mciteSetBstMidEndSepPunct{\mcitedefaultmidpunct}
{\mcitedefaultendpunct}{\mcitedefaultseppunct}\relax
\EndOfBibitem
\bibitem[Aguado \latin{et~al.}(2003)Aguado, Bernasconi, and Madden]{Aguado2003}
Aguado,~A.; Bernasconi,~L.; Madden,~P.~A. Interionic potentials from ab initio
  molecular dynamics: The alkaline earth oxides {CaO}, {SrO}, and {BaO}.
  \emph{The Journal of Chemical Physics} \textbf{2003}, \emph{118},
  5704--5717\relax
\mciteBstWouldAddEndPuncttrue
\mciteSetBstMidEndSepPunct{\mcitedefaultmidpunct}
{\mcitedefaultendpunct}{\mcitedefaultseppunct}\relax
\EndOfBibitem
\bibitem[Alfredsson \latin{et~al.}(2005)Alfredsson, Brodholt, Wilson, Price,
  Cor{\`{a}}, Calleja, Bruin, Blanshard, and Tyer]{Alfredsson2005}
Alfredsson,~M.; Brodholt,~J.~P.; Wilson,~P.~B.; Price,~G.~D.; Cor{\`{a}},~F.;
  Calleja,~M.; Bruin,~R.; Blanshard,~L.~J.; Tyer,~R.~P. Structural and magnetic
  phase transitions in simple oxides using hybrid functionals. \emph{Molecular
  Simulation} \textbf{2005}, \emph{31}, 367--377\relax
\mciteBstWouldAddEndPuncttrue
\mciteSetBstMidEndSepPunct{\mcitedefaultmidpunct}
{\mcitedefaultendpunct}{\mcitedefaultseppunct}\relax
\EndOfBibitem
\bibitem[Cinthia \latin{et~al.}(2015)Cinthia, Priyanga, Rajeswarapalanichamy,
  and Iyakutti]{Cinthia2015}
Cinthia,~A.~J.; Priyanga,~G.~S.; Rajeswarapalanichamy,~R.; Iyakutti,~K.
  Structural, electronic and mechanical properties of alkaline earth metal
  oxides {MO} (M=Be, Mg, Ca, Sr, Ba). \emph{Journal of Physics and Chemistry of
  Solids} \textbf{2015}, \emph{79}, 23--42\relax
\mciteBstWouldAddEndPuncttrue
\mciteSetBstMidEndSepPunct{\mcitedefaultmidpunct}
{\mcitedefaultendpunct}{\mcitedefaultseppunct}\relax
\EndOfBibitem
\bibitem[Pozhivatenko(2020)]{Pozhivatenko2020}
Pozhivatenko,~V. Ionic Character, Phase Transitions, and Metallization in
  Alkaline-Earth Metal Oxides and Chalcogenides under Pressure. \emph{Ukrainian
  Journal of Physics} \textbf{2020}, \emph{65}, 1022\relax
\mciteBstWouldAddEndPuncttrue
\mciteSetBstMidEndSepPunct{\mcitedefaultmidpunct}
{\mcitedefaultendpunct}{\mcitedefaultseppunct}\relax
\EndOfBibitem
\bibitem[Kunduru \latin{et~al.}(2021)Kunduru, Yedukondalu, Roshan, Sripada, and
  Sainath]{Kunduru2021}
Kunduru,~L.; Yedukondalu,~N.; Roshan,~S.~R.; Sripada,~S.; Sainath,~M.
  Structural phase transition and electronic structure of binary {CaO} and
  {SrO} under high pressure. \emph{Materials Today: Proceedings} \textbf{2021},
  \relax
\mciteBstWouldAddEndPunctfalse
\mciteSetBstMidEndSepPunct{\mcitedefaultmidpunct}
{}{\mcitedefaultseppunct}\relax
\EndOfBibitem
\bibitem[K\"{u}rk{\c{c}}\"{u} \latin{et~al.}(2018)K\"{u}rk{\c{c}}\"{u}, Merdan,
  and Yam{\c{c}}{\i}{\c{c}}{\i}er]{Krk2018}
K\"{u}rk{\c{c}}\"{u},~C.; Merdan,~Z.; Yam{\c{c}}{\i}{\c{c}}{\i}er,~{\c{C}}.
  Structural phase transition and electronic properties of {CaO} under high
  pressure. \emph{Materials Research Express} \textbf{2018}, \emph{5},
  125903\relax
\mciteBstWouldAddEndPuncttrue
\mciteSetBstMidEndSepPunct{\mcitedefaultmidpunct}
{\mcitedefaultendpunct}{\mcitedefaultseppunct}\relax
\EndOfBibitem
\bibitem[Kholiya \latin{et~al.}(2010)Kholiya, Verma, Pandey, and
  Gupta]{Kholiya2010}
Kholiya,~K.; Verma,~S.; Pandey,~K.; Gupta,~B. High-pressure behavior of calcium
  chalcogenides. \emph{Physica B: Condensed Matter} \textbf{2010}, \emph{405},
  2683--2686\relax
\mciteBstWouldAddEndPuncttrue
\mciteSetBstMidEndSepPunct{\mcitedefaultmidpunct}
{\mcitedefaultendpunct}{\mcitedefaultseppunct}\relax
\EndOfBibitem
\bibitem[Jeanloz \latin{et~al.}(1979)Jeanloz, Ahrens, Mao, and
  Bell]{Jeanloz1979}
Jeanloz,~R.; Ahrens,~T.~J.; Mao,~H.~K.; Bell,~P.~M. B1-B2 Transition in Calcium
  Oxide from Shock-Wave and Diamond-Cell Experiments. \emph{Science}
  \textbf{1979}, \emph{206}, 829--830\relax
\mciteBstWouldAddEndPuncttrue
\mciteSetBstMidEndSepPunct{\mcitedefaultmidpunct}
{\mcitedefaultendpunct}{\mcitedefaultseppunct}\relax
\EndOfBibitem
\bibitem[Bhardwaj and Singh(2010)Bhardwaj, and Singh]{Bhardwaj2010}
Bhardwaj,~P.; Singh,~S. Role of temperature in the numerical analysis of {CaO}
  under high pressure. \emph{Open Chemistry} \textbf{2010}, \emph{8},
  126--133\relax
\mciteBstWouldAddEndPuncttrue
\mciteSetBstMidEndSepPunct{\mcitedefaultmidpunct}
{\mcitedefaultendpunct}{\mcitedefaultseppunct}\relax
\EndOfBibitem
\bibitem[Hou \latin{et~al.}(2021)Hou, Tan, Hu, Chen, and Geng]{Hou2021}
Hou,~X.-Y.; Tan,~J.; Hu,~C.-E.; Chen,~X.-R.; Geng,~H.-Y. Thermoelectric
  properties of strontium oxide under pressure: First-principles study.
  \emph{Physics Letters A} \textbf{2021}, \emph{390}, 127083\relax
\mciteBstWouldAddEndPuncttrue
\mciteSetBstMidEndSepPunct{\mcitedefaultmidpunct}
{\mcitedefaultendpunct}{\mcitedefaultseppunct}\relax
\EndOfBibitem
\bibitem[Luka{\v{c}}evi{\'{c}}(2011)]{SrO_Lukaevi2011}
Luka{\v{c}}evi{\'{c}},~I. High pressure lattice dynamics, dielectric and
  thermodynamic properties of {SrO}. \emph{Physica B: Condensed Matter}
  \textbf{2011}, \emph{406}, 3410--3416\relax
\mciteBstWouldAddEndPuncttrue
\mciteSetBstMidEndSepPunct{\mcitedefaultmidpunct}
{\mcitedefaultendpunct}{\mcitedefaultseppunct}\relax
\EndOfBibitem
\bibitem[Salam(2019)]{AbdusSalam2019}
Salam,~M. M.~A. First principles study of structural, elastic and electronic
  structural properties of strontium chalcogenides. \emph{Chinese Journal of
  Physics} \textbf{2019}, \emph{57}, 418--434\relax
\mciteBstWouldAddEndPuncttrue
\mciteSetBstMidEndSepPunct{\mcitedefaultmidpunct}
{\mcitedefaultendpunct}{\mcitedefaultseppunct}\relax
\EndOfBibitem
\bibitem[Souadkia \latin{et~al.}(2012)Souadkia, Bennecer, and
  Kalarasse]{Souadkia2012}
Souadkia,~M.; Bennecer,~B.; Kalarasse,~F. Ab initio lattice dynamics and
  thermodynamic properties of {SrO} under pressure. \emph{Journal of Physics
  and Chemistry of Solids} \textbf{2012}, \emph{73}, 129--135\relax
\mciteBstWouldAddEndPuncttrue
\mciteSetBstMidEndSepPunct{\mcitedefaultmidpunct}
{\mcitedefaultendpunct}{\mcitedefaultseppunct}\relax
\EndOfBibitem
\bibitem[Sato and Jeanloz(1981)Sato, and Jeanloz]{Sato1981}
Sato,~Y.; Jeanloz,~R. Phase transition in {SrO}. \emph{Journal of Geophysical
  Research} \textbf{1981}, \emph{86}, 11773\relax
\mciteBstWouldAddEndPuncttrue
\mciteSetBstMidEndSepPunct{\mcitedefaultmidpunct}
{\mcitedefaultendpunct}{\mcitedefaultseppunct}\relax
\EndOfBibitem
\bibitem[Liu and Bassett(1972)Liu, and Bassett]{Liu1972}
Liu,~L.-g.; Bassett,~W.~A. Effect of pressure on the crystal structure and the
  lattice parameters of BaO. \emph{Journal of Geophysical Research (1896-1977)}
  \textbf{1972}, \emph{77}, 4934--4937\relax
\mciteBstWouldAddEndPuncttrue
\mciteSetBstMidEndSepPunct{\mcitedefaultmidpunct}
{\mcitedefaultendpunct}{\mcitedefaultseppunct}\relax
\EndOfBibitem
\bibitem[Weir \latin{et~al.}(1986)Weir, Vohra, and Ruoff]{Weir1986}
Weir,~S.~T.; Vohra,~Y.~K.; Ruoff,~A.~L. High-pressure phase transitions and the
  equations of state of {BaS} and {BaO}. \emph{Physical Review B}
  \textbf{1986}, \emph{33}, 4221--4226\relax
\mciteBstWouldAddEndPuncttrue
\mciteSetBstMidEndSepPunct{\mcitedefaultmidpunct}
{\mcitedefaultendpunct}{\mcitedefaultseppunct}\relax
\EndOfBibitem
\bibitem[Rieder \latin{et~al.}(1973)Rieder, Weinstein, Cardona, and
  Bilz]{Rieder1973}
Rieder,~K.~H.; Weinstein,~B.~A.; Cardona,~M.; Bilz,~H. Measurement and
  Comparative Analysis of the Second-Order Raman Spectra of the Alkaline-Earth
  Oxides with a {NaCl} Structure. \emph{Physical Review B} \textbf{1973},
  \emph{8}, 4780--4786\relax
\mciteBstWouldAddEndPuncttrue
\mciteSetBstMidEndSepPunct{\mcitedefaultmidpunct}
{\mcitedefaultendpunct}{\mcitedefaultseppunct}\relax
\EndOfBibitem
\bibitem[Galtier \latin{et~al.}(1972)Galtier, Montaner, and Vidal]{Galtier1972}
Galtier,~M.; Montaner,~A.; Vidal,~G. Phonons Optiques de {CaO}, {SrO}, {BaO} Au
  Centre de la Zone de Brillouin {\`{a}} 300 et 17K. \emph{Journal of Physics
  and Chemistry of Solids} \textbf{1972}, \emph{33}, 2295--2302\relax
\mciteBstWouldAddEndPuncttrue
\mciteSetBstMidEndSepPunct{\mcitedefaultmidpunct}
{\mcitedefaultendpunct}{\mcitedefaultseppunct}\relax
\EndOfBibitem
\bibitem[Miyanishi \latin{et~al.}(2015)Miyanishi, Tange, Ozaki, Kimura, Sano,
  Sakawa, Tsuchiya, and Kodama]{Miyanishi2015}
Miyanishi,~K.; Tange,~Y.; Ozaki,~N.; Kimura,~T.; Sano,~T.; Sakawa,~Y.;
  Tsuchiya,~T.; Kodama,~R. Laser-shock compression of magnesium oxide in the
  warm-dense-matter regime. \emph{Physical Review E} \textbf{2015},
  \emph{92}\relax
\mciteBstWouldAddEndPuncttrue
\mciteSetBstMidEndSepPunct{\mcitedefaultmidpunct}
{\mcitedefaultendpunct}{\mcitedefaultseppunct}\relax
\EndOfBibitem
\bibitem[Root \latin{et~al.}(2015)Root, Shulenburger, Lemke, Dolan, Mattsson,
  and Desjarlais]{Root2015}
Root,~S.; Shulenburger,~L.; Lemke,~R.~W.; Dolan,~D.~H.; Mattsson,~T.~R.;
  Desjarlais,~M.~P. Shock Response and Phase Transitions of {MgO} at Planetary
  Impact Conditions. \emph{Physical Review Letters} \textbf{2015},
  \emph{115}\relax
\mciteBstWouldAddEndPuncttrue
\mciteSetBstMidEndSepPunct{\mcitedefaultmidpunct}
{\mcitedefaultendpunct}{\mcitedefaultseppunct}\relax
\EndOfBibitem
\bibitem[Duan \latin{et~al.}(2008)Duan, Qin, Tang, and Shi]{Duan2008}
Duan,~Y.; Qin,~L.; Tang,~G.; Shi,~L. First-principles study of ground- and
  metastable-state properties of {XO} (X = Be, Mg, Ca, Sr, Ba, Zn and Cd).
  \emph{The European Physical Journal B} \textbf{2008}, \emph{66},
  201--209\relax
\mciteBstWouldAddEndPuncttrue
\mciteSetBstMidEndSepPunct{\mcitedefaultmidpunct}
{\mcitedefaultendpunct}{\mcitedefaultseppunct}\relax
\EndOfBibitem
\bibitem[Rajput and Roy(2020)Rajput, and Roy]{Rajput2020}
Rajput,~K.; Roy,~D.~R. Structure, stability, electronic and thermoelectric
  properties of strontium chalcogenides. \emph{Physica E: Low-dimensional
  Systems and Nanostructures} \textbf{2020}, \emph{119}, 113965\relax
\mciteBstWouldAddEndPuncttrue
\mciteSetBstMidEndSepPunct{\mcitedefaultmidpunct}
{\mcitedefaultendpunct}{\mcitedefaultseppunct}\relax
\EndOfBibitem
\bibitem[Bouchet \latin{et~al.}(2019)Bouchet, Bottin, Recoules, Remus, Morard,
  Bolis, and Benuzzi-Mounaix]{Bouchet2019}
Bouchet,~J.; Bottin,~F.; Recoules,~V.; Remus,~F.; Morard,~G.; Bolis,~R.~M.;
  Benuzzi-Mounaix,~A. Ab initio calculations of the B1-B2 phase transition in
  {MgO}. \emph{Physical Review B} \textbf{2019}, \emph{99}\relax
\mciteBstWouldAddEndPuncttrue
\mciteSetBstMidEndSepPunct{\mcitedefaultmidpunct}
{\mcitedefaultendpunct}{\mcitedefaultseppunct}\relax
\EndOfBibitem
\bibitem[Yu \latin{et~al.}(2019)Yu, Zhang, Zhang, Wang, and Wu]{Yu2019}
Yu,~J.; Zhang,~M.; Zhang,~Z.; Wang,~S.; Wu,~Y. Hybrid-functional calculations
  of electronic structure and phase stability of {MO} (M = Zn, Cd, Be, Mg, Ca,
  Sr, Ba) and related ternary alloy {MxZn}1-{xO}. \emph{{RSC} Advances}
  \textbf{2019}, \emph{9}, 8507--8514\relax
\mciteBstWouldAddEndPuncttrue
\mciteSetBstMidEndSepPunct{\mcitedefaultmidpunct}
{\mcitedefaultendpunct}{\mcitedefaultseppunct}\relax
\EndOfBibitem
\bibitem[Ghebouli \latin{et~al.}(2010)Ghebouli, Ghebouli, Fatmi, and
  Benkerri]{Ghebouli2010}
Ghebouli,~B.; Ghebouli,~M.; Fatmi,~M.; Benkerri,~M. First-principles
  calculations of structural, elastic, electronic and optical properties of
  {XO} (X=Ca, Sr and Ba) compounds under pressure effect. \emph{Materials
  Science in Semiconductor Processing} \textbf{2010}, \emph{13}, 92--101\relax
\mciteBstWouldAddEndPuncttrue
\mciteSetBstMidEndSepPunct{\mcitedefaultmidpunct}
{\mcitedefaultendpunct}{\mcitedefaultseppunct}\relax
\EndOfBibitem
\bibitem[Baltache \latin{et~al.}(2004)Baltache, Khenata, Sahnoun, Driz, Abbar,
  and Bouhafs]{Baltache2004}
Baltache,~H.; Khenata,~R.; Sahnoun,~M.; Driz,~M.; Abbar,~B.; Bouhafs,~B. Full
  potential calculation of structural, electronic and elastic properties of
  alkaline earth oxides {MgO}, {CaO} and {SrO}. \emph{Physica B: Condensed
  Matter} \textbf{2004}, \emph{344}, 334--342\relax
\mciteBstWouldAddEndPuncttrue
\mciteSetBstMidEndSepPunct{\mcitedefaultmidpunct}
{\mcitedefaultendpunct}{\mcitedefaultseppunct}\relax
\EndOfBibitem
\bibitem[Baumeier \latin{et~al.}(2007)Baumeier, Kr\"{u}ger, and
  Pollmann]{Baumeier2007}
Baumeier,~B.; Kr\"{u}ger,~P.; Pollmann,~J. Bulk and surface electronic
  structures of alkaline-earth metal oxides: Bound surface and image-potential
  states from first principles. \emph{Physical Review B} \textbf{2007},
  \emph{76}\relax
\mciteBstWouldAddEndPuncttrue
\mciteSetBstMidEndSepPunct{\mcitedefaultmidpunct}
{\mcitedefaultendpunct}{\mcitedefaultseppunct}\relax
\EndOfBibitem
\bibitem[Beiranvand(2021)]{Beiranvand2021}
Beiranvand,~R. Hybrid exchange{\textendash}correlation energy functionals for
  accurate prediction of the electronic and optical properties of
  alkaline-earth metal oxides. \emph{Materials Science in Semiconductor
  Processing} \textbf{2021}, \emph{135}, 106092\relax
\mciteBstWouldAddEndPuncttrue
\mciteSetBstMidEndSepPunct{\mcitedefaultmidpunct}
{\mcitedefaultendpunct}{\mcitedefaultseppunct}\relax
\EndOfBibitem
\bibitem[Sobolev \latin{et~al.}(2016)Sobolev, Merzlyakov, and
  Sobolev]{Sobolev2016}
Sobolev,~V.~V.; Merzlyakov,~D.~A.; Sobolev,~V.~V. Optical Properties and
  Electronic Structure of {CaO}. \emph{Journal of Applied Spectroscopy}
  \textbf{2016}, \emph{83}, 567--572\relax
\mciteBstWouldAddEndPuncttrue
\mciteSetBstMidEndSepPunct{\mcitedefaultmidpunct}
{\mcitedefaultendpunct}{\mcitedefaultseppunct}\relax
\EndOfBibitem
\bibitem[Kumar \latin{et~al.}(2013)Kumar, Sharma, and Sharma]{Kumar2013}
Kumar,~R.; Sharma,~B.~K.; Sharma,~G. Electronic Structure and Momentum Density
  of {BaO} and {BaS}. \emph{Advances in Condensed Matter Physics}
  \textbf{2013}, \emph{2013}, 1--6\relax
\mciteBstWouldAddEndPuncttrue
\mciteSetBstMidEndSepPunct{\mcitedefaultmidpunct}
{\mcitedefaultendpunct}{\mcitedefaultseppunct}\relax
\EndOfBibitem
\bibitem[xi~Sun \latin{et~al.}(2021)xi~Sun, yu~Liu, li~Wu, yu~Shi, and mei
  Song]{Sun2021}
xi~Sun,~R.; yu~Liu,~T.; li~Wu,~K.; yu~Shi,~C.; mei Song,~J. First-principles
  study of electronic structure and magnetism in {SrO} crystal contained cation
  defects. \emph{Journal of Magnetism and Magnetic Materials} \textbf{2021},
  \emph{522}, 167524\relax
\mciteBstWouldAddEndPuncttrue
\mciteSetBstMidEndSepPunct{\mcitedefaultmidpunct}
{\mcitedefaultendpunct}{\mcitedefaultseppunct}\relax
\EndOfBibitem
\bibitem[Tsuchiya and Kawamura(2001)Tsuchiya, and Kawamura]{Tsuchiya2001}
Tsuchiya,~T.; Kawamura,~K. Systematics of elasticity:Ab initiostudy in B1-type
  alkaline earth oxides. \emph{The Journal of Chemical Physics} \textbf{2001},
  \emph{114}, 10086--10093\relax
\mciteBstWouldAddEndPuncttrue
\mciteSetBstMidEndSepPunct{\mcitedefaultmidpunct}
{\mcitedefaultendpunct}{\mcitedefaultseppunct}\relax
\EndOfBibitem
\bibitem[Guo \latin{et~al.}(2006)Guo, Cheng, Zhou, Liu, and Yang]{Guo2006}
Guo,~Y.-D.; Cheng,~X.-L.; Zhou,~L.-P.; Liu,~Z.-J.; Yang,~X.-D. First-principles
  calculation of elastic and thermodynamic properties of {MgO} and {SrO} under
  high pressure. \emph{Physica B: Condensed Matter} \textbf{2006}, \emph{373},
  334--340\relax
\mciteBstWouldAddEndPuncttrue
\mciteSetBstMidEndSepPunct{\mcitedefaultmidpunct}
{\mcitedefaultendpunct}{\mcitedefaultseppunct}\relax
\EndOfBibitem
\bibitem[Pandey \latin{et~al.}(2010)Pandey, Gupta, Rathi, and
  Goyal]{Pandey2010}
Pandey,~V.; Gupta,~S.; Rathi,~S.; Goyal,~S. Elastic properties of alkaline
  earth oxides under high pressure. \emph{Phase Transitions} \textbf{2010},
  \emph{83}, 1059--1071\relax
\mciteBstWouldAddEndPuncttrue
\mciteSetBstMidEndSepPunct{\mcitedefaultmidpunct}
{\mcitedefaultendpunct}{\mcitedefaultseppunct}\relax
\EndOfBibitem
\bibitem[Singh \latin{et~al.}(2022)Singh, Singh, Singh, and Shukla]{Singh2022}
Singh,~S.; Singh,~D.; Singh,~N.; Shukla,~M. Study of elastic properties of
  prototype solids under high pressure. \emph{Computational Condensed Matter}
  \textbf{2022}, \emph{30}, e00626\relax
\mciteBstWouldAddEndPuncttrue
\mciteSetBstMidEndSepPunct{\mcitedefaultmidpunct}
{\mcitedefaultendpunct}{\mcitedefaultseppunct}\relax
\EndOfBibitem
\bibitem[Kumar \latin{et~al.}(2021)Kumar, Rajput, and Roy]{Rajput2021}
Kumar,~P.; Rajput,~K.; Roy,~D.~R. Structural, electronic, vibrational,
  mechanical and thermoelectric properties of 2D and bulk BaX (X=O, S, Se and
  Te) series under DFT and BTE framework. \emph{Physica E: Low-dimensional
  Systems and Nanostructures} \textbf{2021}, \emph{127}, 114523\relax
\mciteBstWouldAddEndPuncttrue
\mciteSetBstMidEndSepPunct{\mcitedefaultmidpunct}
{\mcitedefaultendpunct}{\mcitedefaultseppunct}\relax
\EndOfBibitem
\bibitem[Luka{\v{c}}evi{\'{c}}(2011)]{Lukaevi2011-BaO}
Luka{\v{c}}evi{\'{c}},~I. High-pressure lattice dynamics and thermodynamics in
  {BaO}. \emph{physica status solidi (b)} \textbf{2011}, \emph{248},
  1405--1411\relax
\mciteBstWouldAddEndPuncttrue
\mciteSetBstMidEndSepPunct{\mcitedefaultmidpunct}
{\mcitedefaultendpunct}{\mcitedefaultseppunct}\relax
\EndOfBibitem
\bibitem[Uludo{\u{g}}an \latin{et~al.}(2001)Uludo{\u{g}}an, {\c{C}}a{\u{g}}In,
  Strachan, and III]{Uludoan2001}
Uludo{\u{g}}an,~M.; {\c{C}}a{\u{g}}In,~T.; Strachan,~A.; III,~W. A.~G.
  \emph{Journal of Computer-Aided Materials Design} \textbf{2001}, \emph{8},
  193--202\relax
\mciteBstWouldAddEndPuncttrue
\mciteSetBstMidEndSepPunct{\mcitedefaultmidpunct}
{\mcitedefaultendpunct}{\mcitedefaultseppunct}\relax
\EndOfBibitem
\bibitem[Amorim \latin{et~al.}(2006)Amorim, Ver{\'{\i}}ssimo-Alves, and
  Rino]{Amorim2006}
Amorim,~R.~G.; Ver{\'{\i}}ssimo-Alves,~M.; Rino,~J.~P. Energetics of phase
  transitions in {BaO} through {DFT} calculations with norm-conserving
  pseudopotentials: {LDA} vs. {GGA} results. \emph{Computational Materials
  Science} \textbf{2006}, \emph{37}, 349--354\relax
\mciteBstWouldAddEndPuncttrue
\mciteSetBstMidEndSepPunct{\mcitedefaultmidpunct}
{\mcitedefaultendpunct}{\mcitedefaultseppunct}\relax
\EndOfBibitem
\bibitem[Jog \latin{et~al.}(1985)Jog, Singh, and Sanyal]{Jog1985}
Jog,~K.~N.; Singh,~R.~K.; Sanyal,~S.~P. Phase transition and high-pressure
  behavior of divalent metal oxides. \emph{Physical Review B} \textbf{1985},
  \emph{31}, 6047--6057\relax
\mciteBstWouldAddEndPuncttrue
\mciteSetBstMidEndSepPunct{\mcitedefaultmidpunct}
{\mcitedefaultendpunct}{\mcitedefaultseppunct}\relax
\EndOfBibitem
\bibitem[Wu(2002)]{wu2002applications}
Wu,~E.~J. Applications of lattice dynamics theory: calculating vibrational
  entropy in alloys and dielectric losses in ceramics. Ph.D.\ thesis,
  Massachusetts Institute of Technology, 2002\relax
\mciteBstWouldAddEndPuncttrue
\mciteSetBstMidEndSepPunct{\mcitedefaultmidpunct}
{\mcitedefaultendpunct}{\mcitedefaultseppunct}\relax
\EndOfBibitem
\bibitem[Musari and Orukombo(2018)Musari, and Orukombo]{Musari2018}
Musari,~A.~A.; Orukombo,~S.~A. Theoretical study of phonon dispersion, elastic,
  mechanical and thermodynamic properties of barium chalcogenides.
  \emph{International Journal of Modern Physics B} \textbf{2018}, \emph{32},
  1850092\relax
\mciteBstWouldAddEndPuncttrue
\mciteSetBstMidEndSepPunct{\mcitedefaultmidpunct}
{\mcitedefaultendpunct}{\mcitedefaultseppunct}\relax
\EndOfBibitem
\bibitem[Chang \latin{et~al.}(1975)Chang, Tompson, G\"{u}rmen, and
  Muhlestein]{Chang1975}
Chang,~S.; Tompson,~C.; G\"{u}rmen,~E.; Muhlestein,~L. Lattice dynamics of
  {BaO}. \emph{Journal of Physics and Chemistry of Solids} \textbf{1975},
  \emph{36}, 769--773\relax
\mciteBstWouldAddEndPuncttrue
\mciteSetBstMidEndSepPunct{\mcitedefaultmidpunct}
{\mcitedefaultendpunct}{\mcitedefaultseppunct}\relax
\EndOfBibitem
\bibitem[Kresse and Furthm\"{u}ller(1996)Kresse, and
  Furthm\"{u}ller]{Kresse1996}
Kresse,~G.; Furthm\"{u}ller,~J. Efficient iterative schemes forab
  initiototal-energy calculations using a plane-wave basis set. \emph{Physical
  Review B} \textbf{1996}, \emph{54}, 11169--11186\relax
\mciteBstWouldAddEndPuncttrue
\mciteSetBstMidEndSepPunct{\mcitedefaultmidpunct}
{\mcitedefaultendpunct}{\mcitedefaultseppunct}\relax
\EndOfBibitem
\bibitem[Togo and Tanaka(2015)Togo, and Tanaka]{togo2015first}
Togo,~A.; Tanaka,~I. First principles phonon calculations in materials science.
  \emph{Scripta Materialia} \textbf{2015}, \emph{108}, 1--5\relax
\mciteBstWouldAddEndPuncttrue
\mciteSetBstMidEndSepPunct{\mcitedefaultmidpunct}
{\mcitedefaultendpunct}{\mcitedefaultseppunct}\relax
\EndOfBibitem
\bibitem[Clark \latin{et~al.}(2005)Clark, Segall, Pickard, Hasnip, Probert,
  Refson, and Payne]{Clark2005}
Clark,~S.~J.; Segall,~M.~D.; Pickard,~C.~J.; Hasnip,~P.~J.; Probert,~M. I.~J.;
  Refson,~K.; Payne,~M.~C. First principles methods using {CASTEP}.
  \emph{Zeitschrift f\"{u}r Kristallographie - Crystalline Materials}
  \textbf{2005}, \emph{220}, 567--570\relax
\mciteBstWouldAddEndPuncttrue
\mciteSetBstMidEndSepPunct{\mcitedefaultmidpunct}
{\mcitedefaultendpunct}{\mcitedefaultseppunct}\relax
\EndOfBibitem
\bibitem[Momma and Izumi(2008)Momma, and Izumi]{VESTA2008}
Momma,~K.; Izumi,~F. VESTA: a three-dimensional visualization system for
  electronic and structural analysis. \emph{Journal of Applied Crystallography}
  \textbf{2008}, \emph{41}, 653--658\relax
\mciteBstWouldAddEndPuncttrue
\mciteSetBstMidEndSepPunct{\mcitedefaultmidpunct}
{\mcitedefaultendpunct}{\mcitedefaultseppunct}\relax
\EndOfBibitem
\bibitem[Hellman \latin{et~al.}(2011)Hellman, Abrikosov, and
  Simak]{TDEP_PRB_2011}
Hellman,~O.; Abrikosov,~I.~A.; Simak,~S.~I. Lattice Dynamics of Anharmonic
  Solids from First Principles. \emph{Phys. Rev. B} \textbf{2011}, \emph{84},
  180301\relax
\mciteBstWouldAddEndPuncttrue
\mciteSetBstMidEndSepPunct{\mcitedefaultmidpunct}
{\mcitedefaultendpunct}{\mcitedefaultseppunct}\relax
\EndOfBibitem
\bibitem[Liu(1971)]{Liu1971}
Liu,~L. A Dense Modification of BaO and Its Crystal Structure. \emph{Journal of
  Applied Physics} \textbf{1971}, \emph{42}, 3702--3704\relax
\mciteBstWouldAddEndPuncttrue
\mciteSetBstMidEndSepPunct{\mcitedefaultmidpunct}
{\mcitedefaultendpunct}{\mcitedefaultseppunct}\relax
\EndOfBibitem
\bibitem[Catti(2003)]{Catti2003}
Catti,~M. Ab initio predicted metastable {TlI}-like phase in {the B}1to B2
  high-pressure transition of {CaO}. \emph{Physical Review B} \textbf{2003},
  \emph{68}\relax
\mciteBstWouldAddEndPuncttrue
\mciteSetBstMidEndSepPunct{\mcitedefaultmidpunct}
{\mcitedefaultendpunct}{\mcitedefaultseppunct}\relax
\EndOfBibitem
\bibitem[Birch(1947)]{EOS_1947}
Birch,~F. Finite Elastic Strain of Cubic Crystals. \emph{Phys. Rev.}
  \textbf{1947}, \emph{71}, 809--824\relax
\mciteBstWouldAddEndPuncttrue
\mciteSetBstMidEndSepPunct{\mcitedefaultmidpunct}
{\mcitedefaultendpunct}{\mcitedefaultseppunct}\relax
\EndOfBibitem
\bibitem[Cinthia \latin{et~al.}(2015)Cinthia, Priyanga, Rajeswarapalanichamy,
  and Iyakutti]{MO-Elastic-CINTHIA201523}
Cinthia,~A.~J.; Priyanga,~G.~S.; Rajeswarapalanichamy,~R.; Iyakutti,~K.
  Structural, electronic and mechanical properties of alkaline earth metal
  oxides MO (M=Be, Mg, Ca, Sr, Ba). \emph{Journal of Physics and Chemistry of
  Solids} \textbf{2015}, \emph{79}, 23--42\relax
\mciteBstWouldAddEndPuncttrue
\mciteSetBstMidEndSepPunct{\mcitedefaultmidpunct}
{\mcitedefaultendpunct}{\mcitedefaultseppunct}\relax
\EndOfBibitem
\bibitem[Mouhat and Coudert(2014)Mouhat, and Coudert]{Mouhat2014}
Mouhat,~F.; Coudert,~F. m. c.-X. Necessary and sufficient elastic stability
  conditions in various crystal systems. \emph{Phys. Rev. B} \textbf{2014},
  \emph{90}, 224104\relax
\mciteBstWouldAddEndPuncttrue
\mciteSetBstMidEndSepPunct{\mcitedefaultmidpunct}
{\mcitedefaultendpunct}{\mcitedefaultseppunct}\relax
\EndOfBibitem
\bibitem[Rakesh~Roshan \latin{et~al.}(2022)Rakesh~Roshan, Yedukondalu,
  Muthaiah, Lavanya, Anees, Kumar, Rao, Ehm, and Parise]{Roshan2021}
Rakesh~Roshan,~S.~C.; Yedukondalu,~N.; Muthaiah,~R.; Lavanya,~K.; Anees,~P.;
  Kumar,~R.~R.; Rao,~T.~V.; Ehm,~L.; Parise,~J.~B. Anomalous Lattice Thermal
  Conductivity in Rocksalt IIA–VIA Compounds. \emph{ACS Applied Energy
  Materials} \textbf{2022}, \emph{5}, 882--896\relax
\mciteBstWouldAddEndPuncttrue
\mciteSetBstMidEndSepPunct{\mcitedefaultmidpunct}
{\mcitedefaultendpunct}{\mcitedefaultseppunct}\relax
\EndOfBibitem
\bibitem[Efthimiopoulos \latin{et~al.}(2010)Efthimiopoulos, Kunc, Karmakar,
  Syassen, Hanfland, and Vajenine]{Efthimiopoulos2010}
Efthimiopoulos,~I.; Kunc,~K.; Karmakar,~S.; Syassen,~K.; Hanfland,~M.;
  Vajenine,~G. Structural transformation and vibrational properties of
  {BaO}$_2$ at high pressures. \emph{Physical Review B} \textbf{2010},
  \emph{82}\relax
\mciteBstWouldAddEndPuncttrue
\mciteSetBstMidEndSepPunct{\mcitedefaultmidpunct}
{\mcitedefaultendpunct}{\mcitedefaultseppunct}\relax
\EndOfBibitem
\bibitem[Yamaoka \latin{et~al.}(1980)Yamaoka, Shimomura, Nakazawa, and
  Fukunaga]{BaO-Expt}
Yamaoka,~S.; Shimomura,~O.; Nakazawa,~H.; Fukunaga,~O. Pressure-induced phase
  transformation in BaS. \emph{Solid State Communications} \textbf{1980},
  \emph{33}, 87--89\relax
\mciteBstWouldAddEndPuncttrue
\mciteSetBstMidEndSepPunct{\mcitedefaultmidpunct}
{\mcitedefaultendpunct}{\mcitedefaultseppunct}\relax
\EndOfBibitem
\bibitem[Lin \latin{et~al.}(2005)Lin, Gong, and Wu]{BaX-Theory-Lin2005}
Lin,~G.~Q.; Gong,~H.; Wu,~P. Electronic properties of barium chalcogenides from
  first-principles calculations: Tailoring wide-band-gap {II}-{VI}
  semiconductors. \emph{Physical Review B} \textbf{2005}, \emph{71}\relax
\mciteBstWouldAddEndPuncttrue
\mciteSetBstMidEndSepPunct{\mcitedefaultmidpunct}
{\mcitedefaultendpunct}{\mcitedefaultseppunct}\relax
\EndOfBibitem
\bibitem[Vetter and Bartels(1973)Vetter, and Bartels]{Vetter1973}
Vetter,~V.; Bartels,~R. {BaO} single crystal elastic constants and their
  temperature dependence. \emph{Journal of Physics and Chemistry of Solids}
  \textbf{1973}, \emph{34}, 1448--1449\relax
\mciteBstWouldAddEndPuncttrue
\mciteSetBstMidEndSepPunct{\mcitedefaultmidpunct}
{\mcitedefaultendpunct}{\mcitedefaultseppunct}\relax
\EndOfBibitem
\bibitem[Chang and Graham(1977)Chang, and Graham]{Chang1977}
Chang,~Z.; Graham,~E. Elastic properties of oxides in the {NaCl}-structure.
  \emph{Journal of Physics and Chemistry of Solids} \textbf{1977}, \emph{38},
  1355--1362\relax
\mciteBstWouldAddEndPuncttrue
\mciteSetBstMidEndSepPunct{\mcitedefaultmidpunct}
{\mcitedefaultendpunct}{\mcitedefaultseppunct}\relax
\EndOfBibitem
\bibitem[Ghebouli \latin{et~al.}(2011)Ghebouli, Ghebouli, Bouhemadou, Fatmi,
  and Bouamama]{Ghebouli2011}
Ghebouli,~M.; Ghebouli,~B.; Bouhemadou,~A.; Fatmi,~M.; Bouamama,~K. Structural,
  electronic, optical and thermodynamic properties of {Sr$_x$Ca}$_{1-x}$O,
  {Ba$_x$Sr}$_{1-x}$O and {Ba$_x$Ca}$_{1-x}$O alloys. \emph{Journal of Alloys
  and Compounds} \textbf{2011}, \emph{509}, 1440--1447\relax
\mciteBstWouldAddEndPuncttrue
\mciteSetBstMidEndSepPunct{\mcitedefaultmidpunct}
{\mcitedefaultendpunct}{\mcitedefaultseppunct}\relax
\EndOfBibitem
\bibitem[Cortona and Monteleone(1996)Cortona, and Monteleone]{Cortona1996}
Cortona,~P.; Monteleone,~A.~V. Ab initiocalculations of cohesive and structural
  properties of the alkali-earth oxides. \emph{Journal of Physics: Condensed
  Matter} \textbf{1996}, \emph{8}, 8983--8994\relax
\mciteBstWouldAddEndPuncttrue
\mciteSetBstMidEndSepPunct{\mcitedefaultmidpunct}
{\mcitedefaultendpunct}{\mcitedefaultseppunct}\relax
\EndOfBibitem
\end{mcitethebibliography}

\clearpage




\clearpage

\clearpage
\begin{table}[tbp]
\caption{Calculated equilibrium lattice constants (a, c in \AA) of BaO at ambient and high pressure and are compared with the available X-ray diffraction data and other first principles calculations.}
\label{table1}
\begin{tabular}{ccccc} \hline
Phase  &Pressure  &   This work    &  Expt.     &  Others        \\ \hline
 
B1     & 0   &  a=5.533 &  5.520$^a$  &  5.562$^b$, 5.454$^b$ ,5.614$^c$, \\ 
       &     &    -   &       -       &   5.604$^d$, 5.552$^e$,5.462$^f$ \\
B8 &  13.9  & a= 3.612,c= 6.365  & a= 3.617,c= 6.349 $^g$   &- \\ 
d-B2  & 60.5    & a= 4.187,c= 3.007  &a= 4.1,c= 2.998 $^g$   &-  \\ 
B2 & 130   &  a=2.8 & -  & - \\ 
 \hline
\end{tabular}
\\ $^a$Ref.\cite{BaO-Expt} $^b$Ref. \cite{BaX-Theory-Lin2005} $^c$Ref.\cite{Rajput2021} $^d$Ref.\cite{MO-Elastic-CINTHIA201523}
$^e$Ref. \cite{Tsuchiya2001} $^f$ Ref.\cite{Pozhivatenko2020} $^g$ Ref.\cite{Weir1986} 
\end{table}

\begin{table}[tbp]
\caption{Calculated phase transition pressure (P$_t$, in GPa) and volume collapse (VC in $\%$) of BaO and are compared with the available experimental data and other first principles calculations.}
\label{table2}
\begin{tabular}{cccccccc} \hline
Parameter &  Transition          &    This Work   &     Expt.     &  Others        \\ \hline
P$_t$        &  B1 $\rightarrow$ B8 &     5.1        &   9.2$^a$,10 $^b$          &  5 $^c$,4.3 $^d$              \\
          &  B8 $\rightarrow$ d-B2 &   19.5           &   18$^a$,18.8 $^b$           & 13 $^c$,23.2$^d$                \\
VC         &  B1 $\rightarrow$ B8 &    3.65          &    5$^a$           &   3.6$^e$,1.5$^f$               \\
          &  B8 $\rightarrow$ d-B2 &      6.4        &     7$^a$          &       8.7$^e$           \\
           
 \hline
\end{tabular}
\\ $^a$ Ref. \cite{Liu1972} $^b$ Ref.\cite{Weir1986} $^c$ Ref.\cite{Alfredsson2005} $^d$ Ref.\cite{Amorim2006}  $^e$Ref.\cite{Uludoan2001} $^f$Ref.\cite{Lukaevi2011-BaO}
\end{table}

\begin{table}[tbp]
\caption{Calculated equilibrium bulk modulus (B$_0$) and its pressure derivative (B$_0$') of ambient and high pressure phases of BaO and are compared with the available experimental data and previous first principles calculations.}
\label{table3}
\begin{tabular}{cccccccc} \hline
Phase &  Pressure range & B$_0$ (B$_0$')       &  Expt.                                    &  Others        \\ \hline
B1    &  0-9            &  71.4 (4.35)         &  69 $^a$ ,66.2 $^b$ , 61$^c$  & 68.912 $^d$ (4.09$^d$), 73$^e$ (4.3$^e$), \\
      &                 &   -                  &  74.1$^b$, 74.06$^h$ (5.07$^h$)                   & 71.22 $^f$ (4.52$^f$),75.7 $^g$,   \\
      &                 &                      &                                         &  62.75 $^i$, 71.72 $^j$ ,73 $^k$   \\
B8    &  6-30           &  68.6(4.32)          & -  &  68.04$^f$ (4.19$^f$),72.7$^g$  \\  
d-B2  &  20-120         &  51.3 (4.67)         & 33.2 $^b$ (6.02$^b$), &  27.13 $^f$ ,41.8$^g$  \\  
B2    &  70-150         &  90.6 (4.15)         & -                                        & 74.69$^f$ (4.07$^f$), 72.57 $^d$ (3.7$^d$),  \\
      &                 &                      &                                          &  86 $^l$,59.1$^g$  \\  \hline         
\end{tabular}
\\ $^a$ Ref.\cite{Liu1972}  $^b$ Ref.\cite{Weir1986}  $^c$ Ref.\cite{Vetter1973} $^d$ Ref.\cite{Pozhivatenko2020} $^e$Ref. \cite{Tsuchiya2001}  $^f$Ref.\cite{Uludoan2001} $^g$ Ref.\cite{Amorim2006} $^h$Ref.  \cite{Chang1977}  $^i$Ref. \cite{Pandey2010}  $^j$Ref. \cite{Ghebouli2011} $^k$ Ref.\cite{Jog1985}   $^l$ Ref. \cite{Cortona1996}
\end{table}



\end{document}




\clearpage
\begin{figure}
\centering
\includegraphics[width=0.9\columnwidth]{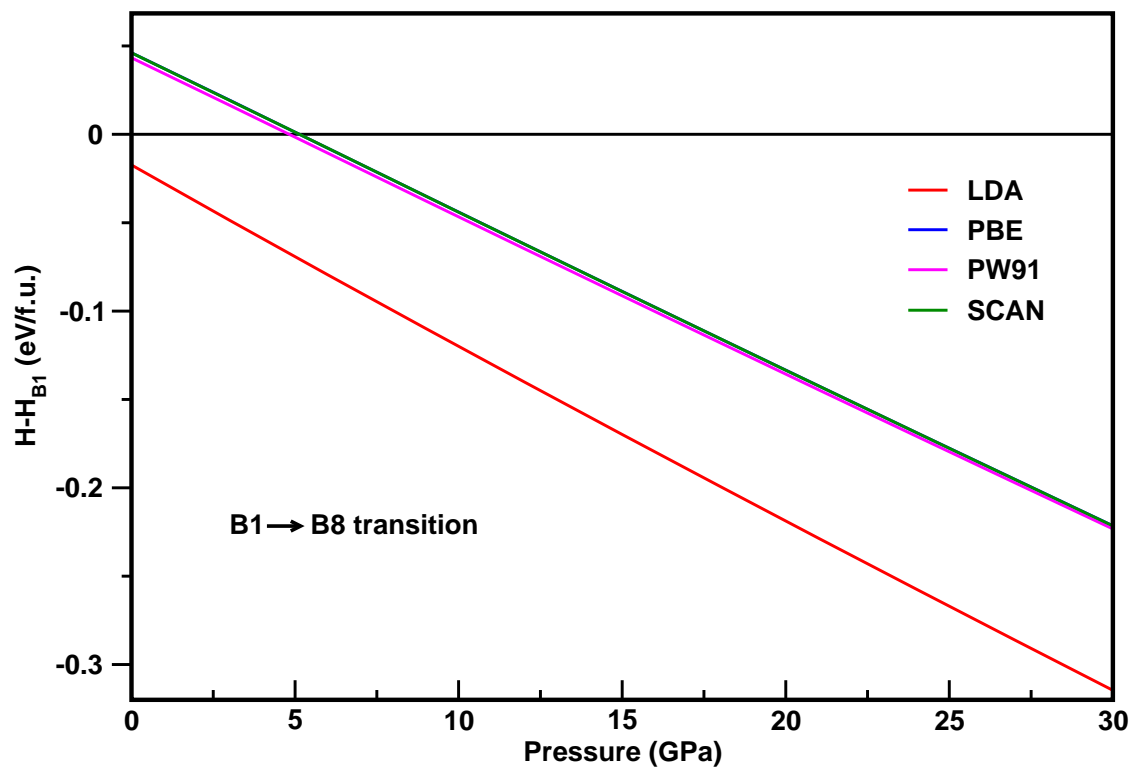}
\caption{Calculated relative enthalpy difference of B8 phase w.r.t B1 phase for BaO using various functionals such as LDA, PBE, PW91 and SCAN.}
\label{str}
\end{figure}

\begin{figure}
\centering
\includegraphics[width=0.9\columnwidth]{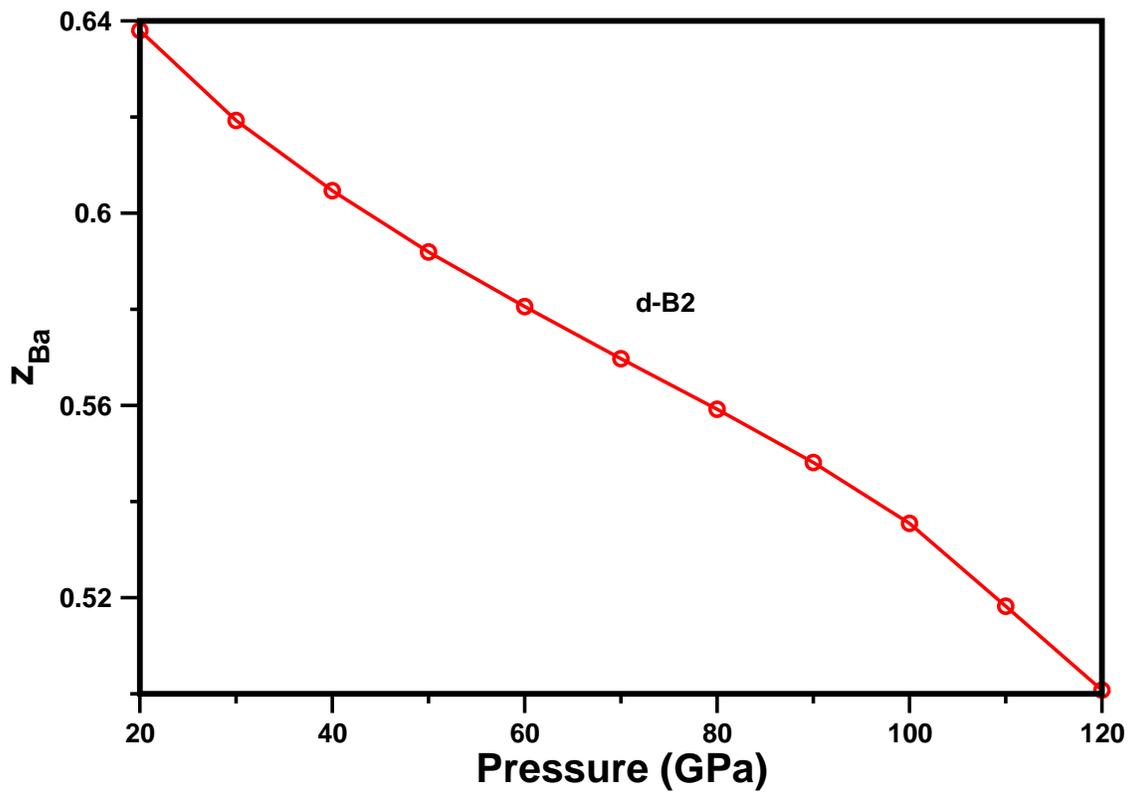}
\caption{Displacive nature of $z_{Ba}$ as a function of pressure for d-B2 phase in the pressure range of 20-120 GPa.}
\label{str}
\end{figure}

\begin{figure}
\centering
\includegraphics[width=0.9\columnwidth]{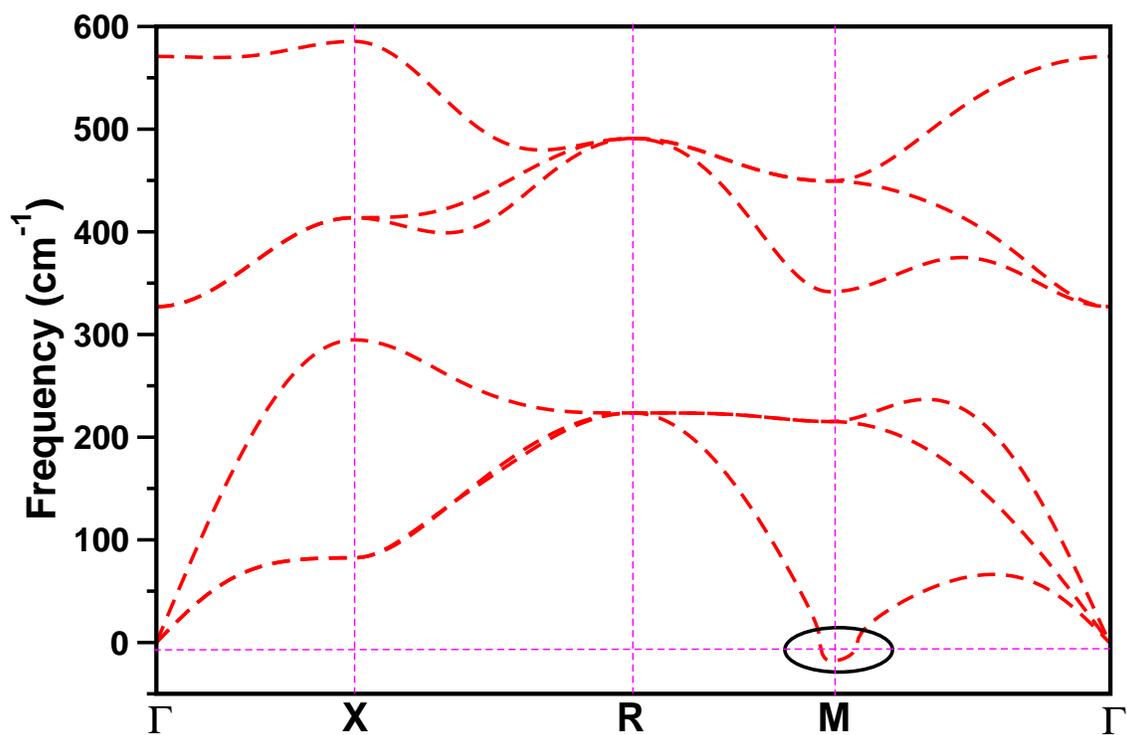}
\caption{Calculated phonon dispersion curves of B2 phase at 80 GPa and 300 K using ab-initio molecular dynamics (AIMD) and temperature dependent effective potential (TDEP).}
\label{str}
\end{figure}

